\documentclass[pdflatex,sn-mathphys-num]{sn-jnl}

\usepackage{times}  
\usepackage{helvet}  
\usepackage{courier}  
\usepackage{graphicx} 
\usepackage{natbib}  
\usepackage{caption} 
\usepackage{newfloat}
\usepackage{listings}
\usepackage{booktabs}
\usepackage{multirow}
\usepackage{xcolor}
\DeclareCaptionStyle{ruled}{labelfont=normalfont,labelsep=colon,strut=off} 
\usepackage{amsmath,amssymb,amsfonts}
\usepackage{graphicx}
\usepackage[utf8]{inputenc} 
\usepackage[T1]{fontenc}    
\usepackage{url}            
\usepackage{booktabs}       
\usepackage{amsfonts}       
\usepackage{nicefrac}       
\usepackage{microtype}      
\usepackage{xcolor}
\usepackage[lined,ruled,vlined,boxed,commentsnumbered]{algorithm2e}
\usepackage{algorithmicx}
\usepackage{algcompatible}
\usepackage{algpseudocode}
\usepackage{subcaption}
\usepackage{enumitem}
\theoremstyle{thmstyleone}%
\newtheorem{theorem}{Theorem}
\theoremstyle{thmstyletwo}%

\theoremstyle{thmstylethree}%

\title{A Bi-directional Multi-solution Scalable Grover Search Algorithm}

\author[1]{\fnm{Debanjan} \sur{Konar}}

\author[1]{\fnm{Zain} \sur{Hafeez}}

\author*[1]{\fnm{Vaneet} \sur{Aggarwal}}\email{vaneet@purdue.edu}

\affil*[1]{\orgname{Purdue University},  \city{West Lafayette}, \postcode{47906}, \state{Indiana}, \country{USA}}

\begin{document}

\abstract{
 Grover's search algorithms, including various Partial Grover Searches (PGS), experience scaling problems when multiple solutions are sought, as the number of iterations increases with the number of solutions or marked states, making implementation more computationally expensive. Inspired by recent PGS algorithms for multi-solution searchers, this article proposes a scalable Grover quantum search algorithm, referred to as Bi-directional Multi-solution scalable Grover Search (BMGS), to efficiently search for an arbitrary number of solutions from an unstructured database. We introduced a novel multi-segment bidirectional search tactic with PGS across multiple equal segments of each state, starting from an initial state and multiple marked states in parallel, obviating the need for merge operations. We have shown in this work that for each solution our novel approach requires at most $\sqrt{\mathcal{N}}\left (1- \sqrt{\frac{1}{b^{\lfloor\frac{r}{dk}\rfloor}}}\right)$ iterations (here, $\mathcal{N}=2^r$ elements, $k=\log_2 b$, $d$ is the number of equal segments on $r$ qubits, and $b$ is the branching factor). Our proposed BMGS algorithm is benchmarked against state-of-the-art Depth First Grover Search (DFGS) and PGS implementations for an arbitrary number of solutions, ranging from $2$ to $20$ qubits, as a proof of concept. We also show that our BMGS requires fewer iterations for shallow quantum circuits and achieves an optimal $\mathcal{O}$($\sqrt{s\mathcal{N}}$) average complexity for $s$ solutions, when $dk < r$. The Qiskit Python implementation of the proposed BMGS algorithm is available on GitHub\footnote{\url{https://anonymous.4open.science/r/Multi-Solution-DFGS-BMGS-B507/}}.
}

\maketitle

\section{Introduction}
\label{chapter:intro}

Grover's Search (GS) algorithm~\cite{grover1997} is a significant advancement in quantum computing, revolutionizing database searches by enabling quantum computers to outperform classical computers quadratically in terms of time efficiency. This algorithm, designed for an Oracle model (black box), has been implemented across various platforms~\cite{durr1998, brassard1998, hoyer1998, cerf2000, mahmud2022} by researchers, showcasing its practicality and theoretical efficiency~\cite{zalka1999}. 
GS technique is known for its ability to locate an item in an unstructured list with high probability using only $\mathcal{O}$($\sqrt{N}$)~\cite{grover1997, charles1997} evaluations, a substantial improvement over classical systems. In a $r$-qubit network, Grover's algorithm asserts that there are ${2^r}$ = $\mathcal{N}$ states, each state having an amplitude of $\frac{1}{\sqrt{N}}$ and a chance of $\frac{1}{N}$ of being discovered~\cite{morales2018}. Bennett \emph{et al.}~\cite{charles1997} have emphasized that no quantum approach can asymptotically solve the database search problem in fewer steps than $\mathcal{O}$($\sqrt{\mathcal{N}}$).\\
However, the standard GS algorithm faces challenges with multi-solution search~\cite{nielsen2001}, particularly as the number of qubits or circuit depth increases, leading to high computational time requirements despite its high success rate~\cite{korepin2009, roland2002}. This limitation underscores the need for further advances in quantum search algorithms to efficiently address multi-solution search~\cite{walther2005,boyer1999,buhrman1999,zhang2018}. To address this issue, Quantum Partial Search Algorithms (QPSA) have been developed, focusing on searching for a block or collection of items containing the marked state rather than the state itself to reduce computational complexity~\cite{wang2023,korepin2006}. Grover and Radhakrishnan (GRK) introduced a fast algorithm for Partial Grover Search (PGS), combining local and global searches to achieve $\frac{\pi}{4}\sqrt{\mathcal{N}}\sqrt{1-\frac{1}{b}}$ iterations~\cite{grover2005}, where $b$ is the branching factor, and achieve an upper bound of $\mathcal{O}$($\sqrt {\frac{\mathcal{N}}{s}}$) for each solution in multi-solution searches, where $s$ is the number of solutions. However, despite the efficient handling of a multi-solution search, certain issues persist in these partial search algorithms, such as the possibility of receiving repeats and the lack of a termination condition for an unspecified number of solutions~\cite{wang2023}. Recently, Guo~\cite{guo2022} proposed an improved version of QPSA, known as the Depth-First Grover Search (DFGS) algorithm, using the Depth-First Search (DFS) and hybrid quantum-classical algorithms for an arbitrary number of solutions. However, DFGS suffers from a complex amplitude interception-based search and a higher number of Oracle calls $\sqrt{\mathcal{N}}(1-\sqrt{\frac{1}{b^r}})$ (where $r = \log_2 \mathcal{N}$) with the time complexity of $\mathcal{O}$($s\sqrt{\mathcal{N}}$), which is not optimal.\\
Inspired by previous work on QPSA and DFGS, this work introduces a Bi-directional Multi-solution scalable Grover Search (BMGS) algorithm, which has a number of Oracle calls as $\sqrt{\mathcal{N}}\left (1- \sqrt{\frac{1}{b^{\lfloor\frac{r}{dk}\rfloor}}}\right)$  to ensure that segmentation does not exceed efficient computational limits. Unlike the DFGS algorithm, our novel BMGS algorithm does not require an additional unitary gate for amplitude interception~\cite{zhang2022}. We introduced a novel multi-segment bidirectional search tactic with PGS across multiple equal segments of each state, starting from an initial state and multiple marked states in parallel, without any additional segment merge operations. The proposed BMGS algorithm showcases scalability for an arbitrary number of solutions, setting it apart from existing QPSA frameworks. The significant contributions outlined in the study are as follows:
\begin{enumerate}[leftmargin=*]
\item The research introduces a scalable BMGS algorithm that leverages PGS at each equal intercept of the search space for an arbitrary number of solutions. This algorithm operates within a hybrid quantum-classical framework, implementing circuits ranging from $2$ to $20$ qubits.
\item Furthermore, an improved version of DFGS~\cite{guo2022}, which eliminates the need for amplitude interception, has been proposed and implemented in a quantum simulation for the first time.
\item In terms of parallelization, our novel BMGS requires iterations as $\sqrt{\mathcal{N}}\left (1- \sqrt{\frac{1}{b^{\lfloor\frac{r}{dk}\rfloor}}}\right)$ (where $k= \lceil\log_2b \rceil$, $b$ is the branching factor, and $dk< r$) and achieves an average optimal complexity of $\mathcal{O}$($\sqrt{s\mathcal{N}}$)  for multi-solution search. Empirically, it has been demonstrated to be a scalable quantum search algorithm and is fastest for shallow quantum circuits, compared to DFGS and PGS, for an arbitrary number of solutions, to the best of our knowledge.
\end{enumerate}

\section{Related Works}
\label{lit:review}

Numerous Grover Search (GS) algorithms have been investigated in the last few decades, including applications in a variety of problem domains~\cite{durr1998, brassard1998, hoyer1998, cerf2000, mahmud2022}, as well as theoretical evidence of the algorithm's efficiency~\cite{zalka1999}. 
Notably, adiabatic quantum computing settings are achieved by extending Grover's quantum search algorithm~\cite{roland2002}. 
Heiligman described a partial quantum search algorithm to find matches between two databases in $\mathcal{O}(\mathcal{N}^{\frac{3}{4}} log\mathcal{N})$~\cite{heiligman2000}. Of late, a quantum pattern matching method, as presented by Niroula \emph{et al.}, matches a search string (pattern) of length $S$ within a larger text of length $T$ with time complexity $\mathcal{O}(\sqrt{S})$~\cite{niroula2021}.  ~\cite{mateus2005} introduces a quantum search method to match the nearest pattern in $\mathcal{O}(\sqrt{S})$, where $S$ is the length of the string, is introduced by \cite{mateus2005}. Tulsi introduced a quantum search algorithm, which finds a single element in common between two sets in $\mathcal{O}(\sqrt{\mathcal{N}})$~\cite{tulsi2012}. Despite improvements in reducing iteration counts, their complexity and computational demands have limited their widespread implementation~\cite{korepin2006}, particularly on Noise-Intermediate-Scale-Quantum (NISQ) computers~\cite{liu2021}.\\
Moreover, standard GS is not suitable for multi-solution searching because it is computationally expensive as the number of qubits or circuit depth increases~\cite{nielsen2001}. Although~\cite{walther2005,boyer1999,buhrman1999,zhang2018} offered some improvements, the following problems persist: (1) it can be possible to receive repeats; (2) the program lacks a termination condition for an unspecified number of solutions, and the computational complexity remains $\mathcal{O}$($\sqrt{\mathcal{N}}$). Of late,  \cite{guo2022} introduced the Depth-First Grover Search Algorithm (DFGS) as an innovative approach to address multi-solution searching problems in unstructured databases with an unknown number of solutions. This algorithm combines Depth-First Search and the Partial Grover Search (PGS) to offer a novel solution. However, the proposed DFGS suffers from complex formulations and a relatively higher number of Oracle calls, $\sqrt{\mathcal{N}}\left(1-\sqrt{\frac{1}{b^r}}\right)$, and is therefore not scalable for wider applications. This motivates us to reduce the number of Oracle calls to $\sqrt{\mathcal{N}}\left (1- \sqrt{\frac{1}{b^{\lfloor\frac{r}{dk}\rfloor}}}\right)$ by introducing a Bi-directional Multi-solution Grover Search (BMGS) with PGS on multiple equal segments of each state, starting from an initial state and multiple marked states in parallel. However, when considering the parallel execution of multiple GS instances ($d$ instances), merging the solutions from these runs typically requires exponential time. In contrast, our BMGS algorithm offers a novel approach in which each segment is processed depth-first, eliminating the need for a merging step and enabling direct attainment of the final solution. 

\subsection{Motivation} 
\label{sec:motiv}

The development of multi-solution scalable Grover Search algorithms is motivated by the need for more efficient and scalable quantum search techniques that can handle real-world applications~\cite{wiebe2015, shen2012}, including database searches, pattern recognition, and various optimization problems~\cite{wang2014, gilliam2021} where multiple satisfactory solutions are common, mitigate the limitations of classical search methods, post-quantum cryptography~\cite{grassl2016} and improve the feasibility of practical quantum computing implementations~\cite{gottesman1997}. 
\begin{algorithm}[th!]
\SetKwBlock{Begin}{Begin}{End}
\SetKwFor{For}{for}{}{end}
\SetAlgoLined
\DontPrintSemicolon
\SetNoFillComment
\LinesNumbered
\KwData{Quantum register $\mathcal{Q}$, and multiple target elements $x$}
\KwResult{Index of list of targets $x$ if found, else $-1$}
\textbf{Procedure BMGS}($k$, $b$, $r$, $x$, \texttt{num\_solutions}, $d$, $A$)\;
\textbf{Initialization of Input Qubits:} \\
Initialize quantum register $\mathcal{Q}$ with $r$ qubits, set to $|0\rangle^r$ and $\mathcal{N} = 2^r$\;
Apply Hadamard gates to all input qubits $\mathcal{Q}$: $H^{\otimes r}|0^r\rangle$\;
\textbf{Initialization of Auxiliary Qubits:} \\
Initialize $k = \lceil \log_2 b \rceil$\;
Initialize auxiliary qubits $A$ to $|0\rangle^d$\;
Set $S_d = \lfloor r / d \rfloor$\;
Set $L = \lfloor S_d / k \rfloor$ 
\ForAll{$\ell \in \{0, 1, \dots, d - 1\}$ \textbf{ in parallel}}{
    \For{$\lambda \gets 0$ \KwTo $L - 1$}{
        \uIf{($\ell \bmod 2 = 0$) \textbf{and} $(\ell \neq d - 1)$}{
            \tcp{Forward Grover search on segment $\ell$}
            \For{$j \gets \ell \cdot S_d$ \KwTo $(\ell+1) \cdot S_d - k$ \textbf{increment by $k$}}{
                $q \gets [j, j + k - 1]$\;
                \If{length($q$) == 1}{
                    HSQ($Q$, $q[0]$, $x[\ell][q[0]]$)\tcp*{Single-qubit search}
                }
                \ElseIf{$\ell = 0$ \textbf{or} $x[\ell][q[0]:q[1]+1] \neq x[\ell-1][q[0]:q[1]+1]$}{
                    PGS($Q$, $q$, $x[\ell][q[0]:q[1]+1]$, $|A\rangle$)\;
                }
            }
        }
        \Else{
            \tcp{Backward Grover search on segment $\ell$}
            \For{$j \gets (\ell+1) \cdot S_d - 1$ \KwTo $\ell \cdot S_d$ \textbf{increment by $-k$}}{
                $q \gets [\max(j - k + 1, \ell \cdot S_d), j]$\;
                \If{length($q$) == 1}{
                    HSQ($\mathcal{Q}$, $q[0]$, $x[\ell][q[0]]$)\;
                }
                \ElseIf{$\ell = 0$ \textbf{or} $x[\ell][q[0]:q[1]+1] \neq x[\ell-1][q[0]:q[1]+1]$}{
                    PGS($\mathcal{Q}$, $q$, $x[\ell][q[0]:q[1]+1]$, $|A\rangle$)\;
                }
            }
        }
    }
}
Measure all qubits in $\mathcal{Q}$\;
\caption{Bi-directional Multi-solution Grover Search Algorithm}
\label{algo:bmgs}
\end{algorithm}

\section{Bi-directional Multi-solution Grover Search}
\label{sec:BMGS}
\begin{figure}[t]
\vspace{-.1in}
\begin{center}
 \subcaptionbox{}{\includegraphics[clip, trim=0.0cm 0.0cm 0.0cm 0.0cm, width=0.8\textwidth]{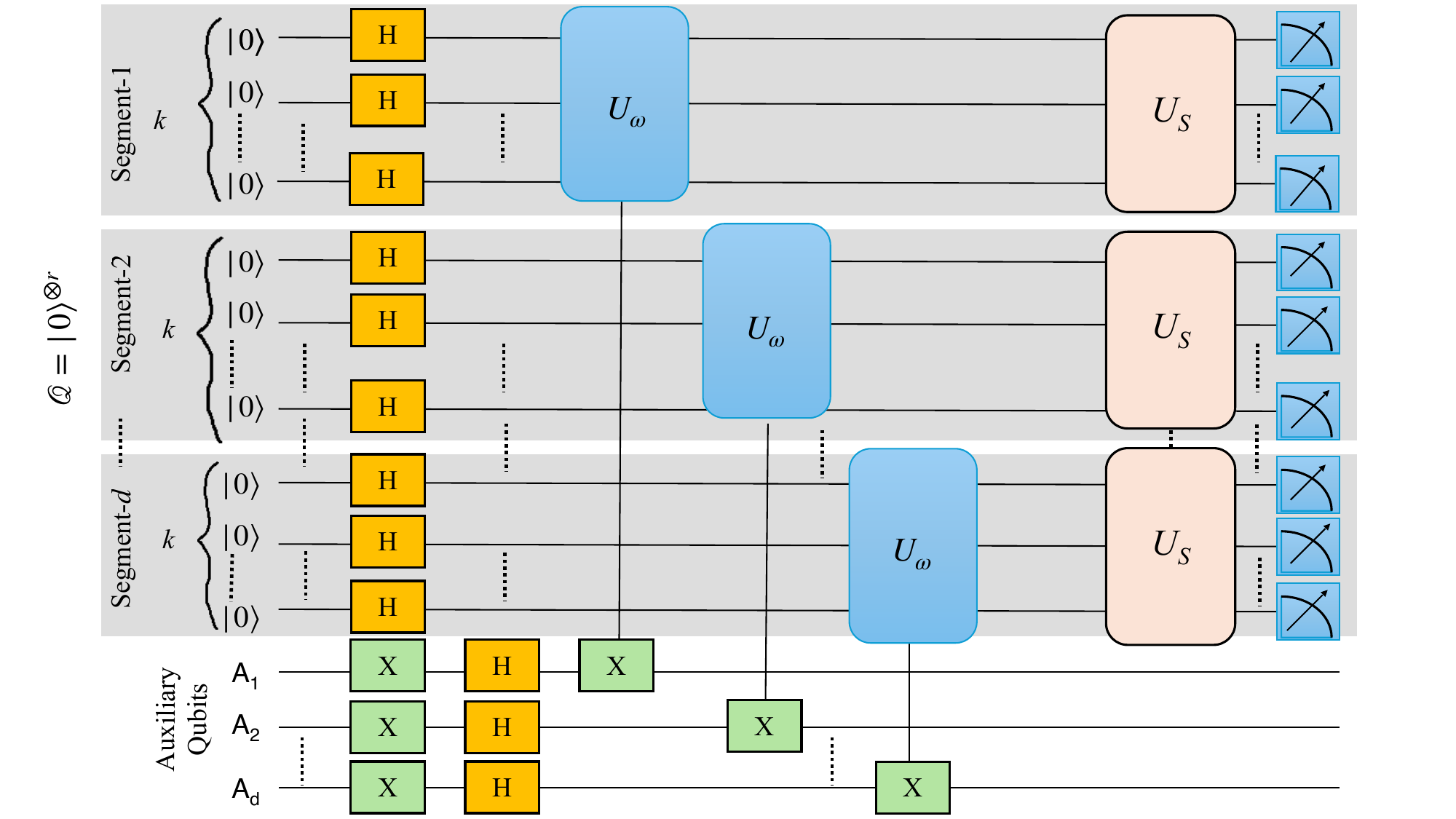}}
\subcaptionbox{}{\includegraphics[width=.6\textwidth]{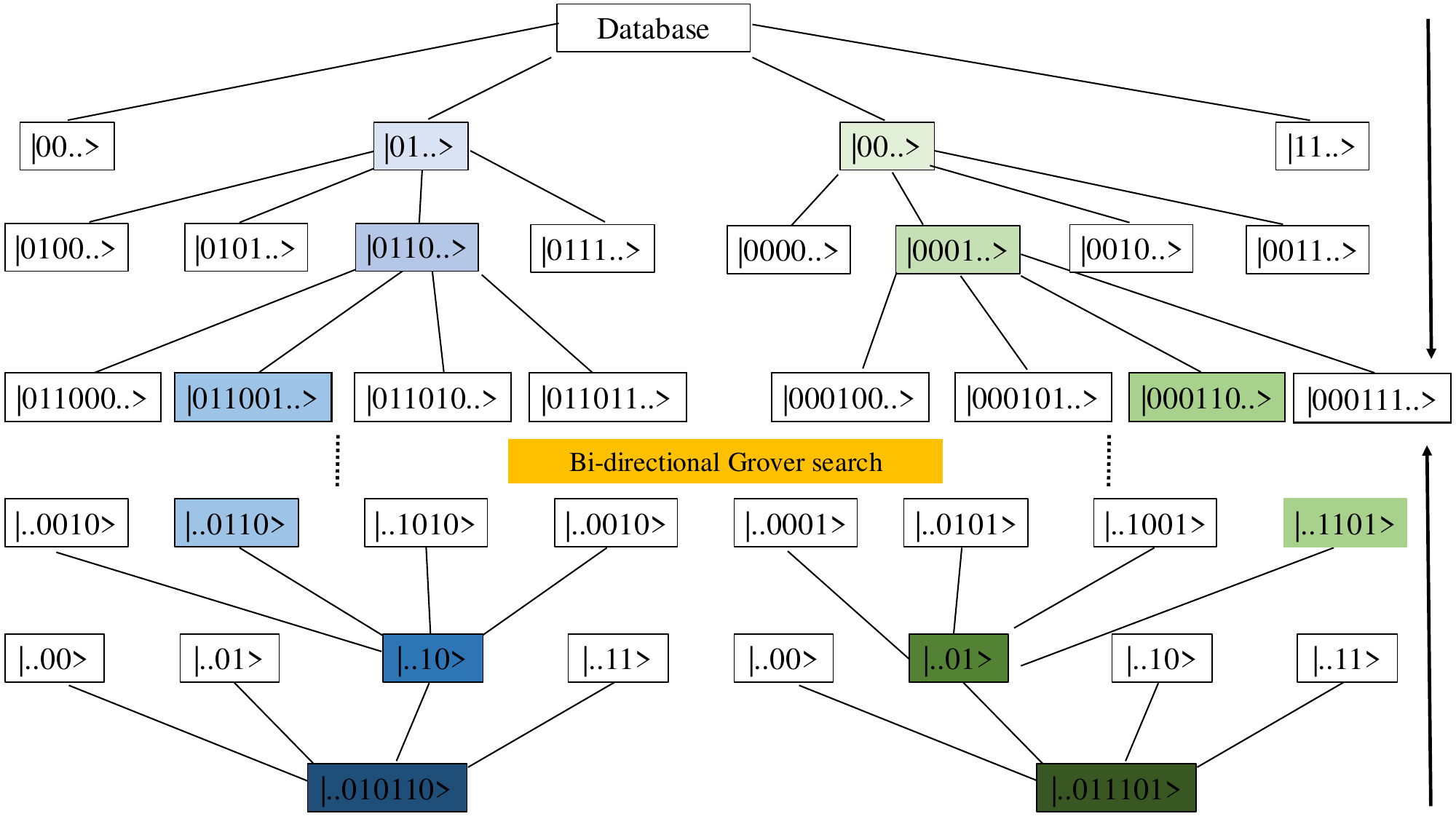}}
\caption{(a) \textcolor{black}{Quantum circuit schematic of the proposed BMGS algorithm. The main quantum register $\mathcal{Q}$ is initialized as $|0\rangle^{\otimes r}$ and transformed into an equal superposition using Hadamard gates ($H$). The $r$-qubit search space is divided into $d$ equal segments, where Partial Grover Search is applied in a bi-directional manner using segmented oracle operators $U_{\omega}$ and diffusion operators $U_{S}$. Auxiliary qubits $A_1, A_2, \dots, A_d$ are used to store intermediate segment-wise search information and consistency checks during forward and backward traversal.} (b) \textcolor{black}{Example of BMGS with branching factor $b=4$, where the database is recursively divided into four subspaces at each layer, and Partial Grover Search is used to determine the next two bits of the solution address through forward and backward search.}}
\label{fig:bdgs}
\end{center}
\end{figure}

 A novel scalable approach combining Bi-Directional Search (BDS)~\cite{korf2003} and Partial Grover Search (PGS)~\cite{grover2005} algorithms is proposed for multi-target items and referred to as Bi-directional Multi-solution Grover Search (BMGS). It relies on a PGS across multiple segments of search states, in a bi-directional fashion, to handle multi-solution search problems with an arbitrary number of solutions. \\
 In the proposed quantum circuit for BMGS, as shown in Figure~\ref{fig:bdgs} (a), $U_\omega$ and $U_s$ are two functions as Oracle and Diffuser, respectively. The main idea of BMGS is to perform forward and backward passes from the initial and multiple target states simultaneously until the two search frontiers intersect at multiple equal intercepts of the search states. The entire solution path is then formed by concatenating the path from the start state with the inverse path from the target state for each multiple intercept. An optimal number of Oracle calls is guaranteed for arbitrary numbers of solutions using the proposed BMGS quantum search algorithm with a time complexity of $\mathcal{O}$($\sqrt{s\mathcal{N}}$) for $s\leqslant d$ since each search only needs to go down to the length of each segment of the depth of the solution in parallel.\\
In this proposed BMGS algorithm with inputs $k$: Number of bits representing segments of the search space, $b$: Branching factor of the bi-directional tree search space, $r$: Total number of qubits representing the entire database, $x$: List of target elements to find, $d$: The number of segments, $s$: Total number of solutions, $\mathcal{A}$: List of auxiliary qubits ($|\mathcal{A}| = d$):
\begin{enumerate}[leftmargin=*]
\item Initially, intercepting each search state of $r$-qubits into $d$-equal parts, each segment comprises at least ($\lfloor \frac{r}{d}\rfloor$) qubits. Applying the PGS on each segment's $\lfloor \frac{r}{d}\ rfloor$ qubits in parallel in a bi-directional fashion for finding the next bits as demonstrated in Equations~\ref{eq:forward} and~\ref{eq:backward}. The remaining qubits (if any ($r-d*\lfloor \frac{r}{d}\rfloor$)) are taken care of using a forward or backward search.
\item We recursively traverse each possible solution interval at each layer using PGS in a bi-directional fashion (quantum operator $\mathcal{I}_G$) for $b$ branches to identify which intervals among the $b$ (branching factor) equal-size split sub-intervals have a solution.
\item Finally, a standard Grover search (using the quantum operator $\mathcal{I}_f$) is involved in determining the precise address of the solution for intervals whose widths are less than or equal to $b$.
\end{enumerate}
A pseudo-code for our BMGS algorithm \textbf{Procedure BMGS()} has been provided in Algorithm 1. A detailed explanation of the algorithm has been provided in \emph{Appendix}.
\begin{algorithm}[htbp]
\SetKwBlock{Begin}{Begin}{End}
\SetKwFor{For}{for}{}{end}
\SetAlgoLined
\DontPrintSemicolon
\SetNoFillComment
\LinesNumbered
  \textbf{Procedure HSQ($\mathcal{Q}, q\left[0\right], x\left[l\right]\left[q\left[0\right]\right]$)}\\
 Apply H gate to qubit $q[0]$\\
 \textbf{if} $x[l][q[0]] == '1'$:\\
 \  Apply $Z$ gate to qubit $q[0]$\\
  Apply H gate to qubit $q[0]$\\
 \textbf{Procedure PGS ($\mathcal{Q}, q, x, \left[l\right]\left[q\left[0\right]:q\left[1\right]+1\right],|A\rangle$)}\\
 Apply Oracle to mark the state corresponding to $x$\\
 Apply the Diffusion operator to amplify the marked state\\
\caption{Procedures HSQ and PGS}
\end{algorithm}

The following expression shows the equal segmentation of $\lfloor \frac{r}{d}\rfloor$ qubits, and PGS are applied on each segment in a bi-directional manner as follows:
\begin{eqnarray}
&|\underbrace{X_0\cdots X_{(\lfloor \frac{r}{d}\rfloor-1)}}_{Forward}^{\xrightarrow{\text{Partial Grover Search}}}
\underbrace{X_{(\lfloor \frac{r}{d}\rceil)}\cdots X_{(2\lfloor \frac{r}{d}\rfloor-1)}}_{Backward}^{\xleftarrow{\text{Partial Grover Search}}}\cdots\nonumber \\ &\underbrace{X_{(d-2)\lfloor \frac{r}{d}\rfloor}\cdots X_{(d-1)\lfloor \frac{r}{d}\rfloor +1}}_{Forward}^{\xrightarrow{\text{Partial Grover Search}}}
\underbrace{X_{(d-1)\lfloor \frac{r}{d}\rfloor+1}\cdots X_{(r-1)}}_{Backward}^{\xleftarrow{\text{Partial Grover Search}}}\rangle
\end{eqnarray}
The proposed BMGS algorithm incorporates a PGS~\cite{korf2003} to determine the next $k= \lceil\log_2 b \rceil$ bits in the forward and backward search in BMGS as follows:
\begin{eqnarray}
&  |\underbrace{X_0X_1\cdots X_iX_{i+1}}_{Found}\underbrace{\square\square}_{Next~k}\cdots\rangle \xrightarrow[Forward]{\text{Partial Grover Search}}\nonumber\\
&|\underbrace{X_0X_1\cdots X_iX_{i+1}}_{Found}\underbrace{X_{i+2}X_{i+3}}_{Next~k}\cdots\rangle\ .
\label{eq:forward}
\end{eqnarray}
\begin{eqnarray}  &|\cdots\underbrace{\square\square}_{Prev.~k}\underbrace{X_i X_{i+1}\cdots X_{r-1} X_r}_{Found}\rangle \xrightarrow[Backward]{\text{Partial Grover Search}}\nonumber\\
&|\cdots\underbrace{X_{i-2} X_{i-1}}_{Prev.~k} \underbrace{X_i X_{i+1}\cdots X_{r-1} X_r}_{Found}\rangle\ .
\label{eq:backward}
\end{eqnarray}
Here, $X_i \in \{0, 1\}$. Two auxiliary qubits are placed to store the following: (1) the bits' found value; and (2) the bits' status (checked and unchecked). The qubits will be initialized into a superposition of this encoding. Assume $\mathcal{N} = 2^r$
\begin{eqnarray}
&|\underbrace{\cdots\cdots}_{Found~l}\rangle \otimes \frac{1}{\sqrt{2^{\lfloor \frac{r}{d} \rfloor-l-1}}} \bigoplus_{i=0}^{\lfloor \frac{r}{d} \rfloor-l-1} (|0\rangle + |1\rangle)\ ~\text{and}\nonumber\\
&|\underbrace{\cdots\cdots}_{Found~l}\rangle \otimes \frac{1}{\sqrt{2^{\lfloor \frac{r}{d} \rfloor-l}}} \bigoplus_{i=r-1}^{\lfloor \frac{r}{d} \rfloor-l} (|0\rangle + |1\rangle)\ .
\end{eqnarray}
As shown in Figure~\ref{fig:bdgs} (a), since the first four bits are determined on the second layer, the first four bits are encoded by the auxiliary qubits, which have the values $A_1$ = $0110\cdots$ and $A_2$ = $1111\cdots$ as follows.
\begin{equation}
\begin{split}
|\underbrace{0110}\rangle \otimes \frac{1}{\sqrt{2^{\lfloor \frac{r}{d} \rfloor-4-1}}} \bigoplus_{i=0}^{\lfloor \frac{r}{d} \rfloor-4-1} (|0\rangle + |1\rangle)\ .
\end{split}
\end{equation}
We aim to find the next $k$ bits using PGS, but the determined bits are fixed. A description of standard GS and PGS is provided in \emph{Appendix}.

\subsection{Computational Complexity Analysis of BMGS}  
\label{iteration:bdgs}

In bi-directional search, if forward ($\mathcal{F}_i$) and backward ($\mathcal{R}_i$) reaches the satisfaction criteria $\mathcal{C} =\prod_{i=1}^{\frac{d}{2}}(\mathcal{F}_i \wedge \mathcal{R}_i)$ for every segment $i$, it is considered that the target item has been found. We can create a database with all $r$-length pathways ($2^r = \mathcal{N}$), $\lambda_0-\lambda_1-\cdots \lambda_{r-1}$. The subpath $\mathcal{F}=\lambda_0-\lambda_1-\cdots \lambda_{\lfloor \frac{r}{dk}\rfloor-1}$ is applied to each vertex once. The proposed BMGS algorithm's initial step aims to increase the amplitude of the target blocks using PGS iteratively. To do it, 
$\sin [(2t +1)\omega] =1$, \emph{i.e.}, $(2t +1)\omega =\frac{\pi}{2}$.
Hence, the maximum number of $\mathcal{I_G}$ iterations at level $\lambda$ of the bi-directional tree search with the $\mu_\lambda$ number of PGS applied on each segment is evaluated as~\cite{grover2005}:
\begin{equation}
\mathcal{I_G}^\lambda=\mu_\lambda\left (\sqrt{\frac{\mathcal{N}}{b^\lambda}}-\sqrt{\frac{\mathcal{N}}{b^{\lambda +1}}}\right).
\end{equation}
Since $(2t +1)\omega =\frac{\pi}{2}$, $\cos[(2t +1)\omega] \approx 0$, which implies that regardless of the number of iterations, the amplitudes of non-target blocks are all set to $0$ 
and $\mathcal{N} = 2^r$. For example, as shown in Figure~\ref{fig:bdgs}, the database requires representing $8$ qubits ($\mathcal{N}=2^r$). $\mathcal{F}$ is the initial $2$ qubit satisfaction condition and $\mathcal{R}$ is the final $2$ qubits' satisfaction requirement for bi-directional search ($d=2$), with $\mathcal{C} =\mathcal{F} \wedge \mathcal{R}$. Therefore, $b = 2^2$.\\ 
According to GRK's PGS~\cite{grover2005}, the total number of queries required for BMGS for a single partial Grover search in both forward and backward passes for each segment of $d$ at level $\lambda$, assuming that, on average, each pass comprises a maximum of $\lfloor \frac{r}{d}\rfloor$ qubits, is 
\vspace{-0.2cm}
\begin{equation}
 \mathcal{I}_G^\lambda =\sqrt{\mathcal{N}}\left (\sqrt{\frac{1}{b^\lambda}}-\sqrt{\frac{1}{b^{\lambda +1}}}\right) \ .  
\end{equation}
\begin{figure*}[thbp]
\centering
\includegraphics[clip, trim=0.0cm 0.0cm 0.0cm 0.0cm, width=0.46\textwidth]{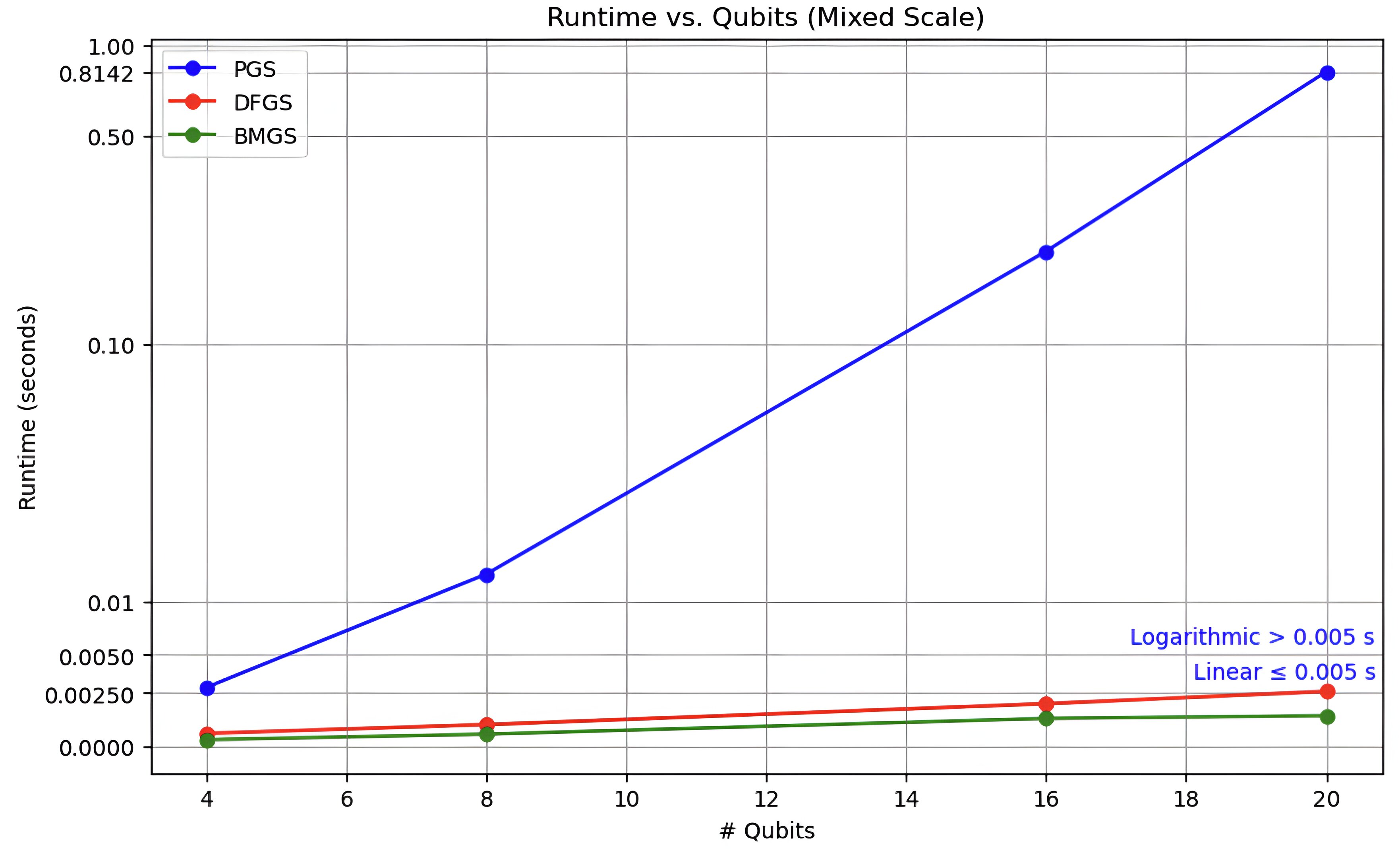}
\includegraphics[clip, trim=0.0cm 0.0cm 0.0cm 0.0cm, width=0.46\textwidth]{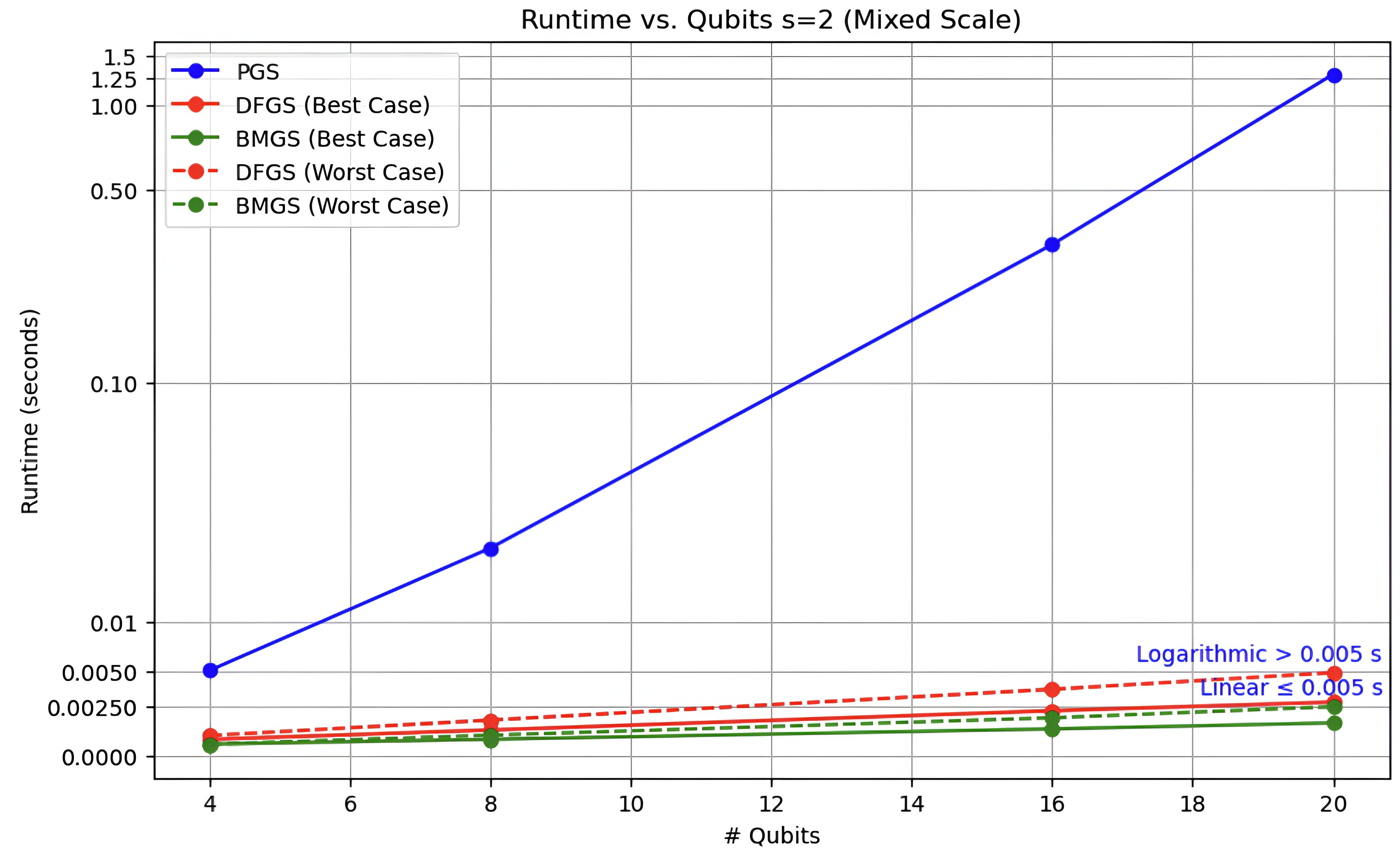}
\includegraphics[clip, trim=0.0cm 0.0cm 0.0cm 0.0cm, width=0.46\textwidth]{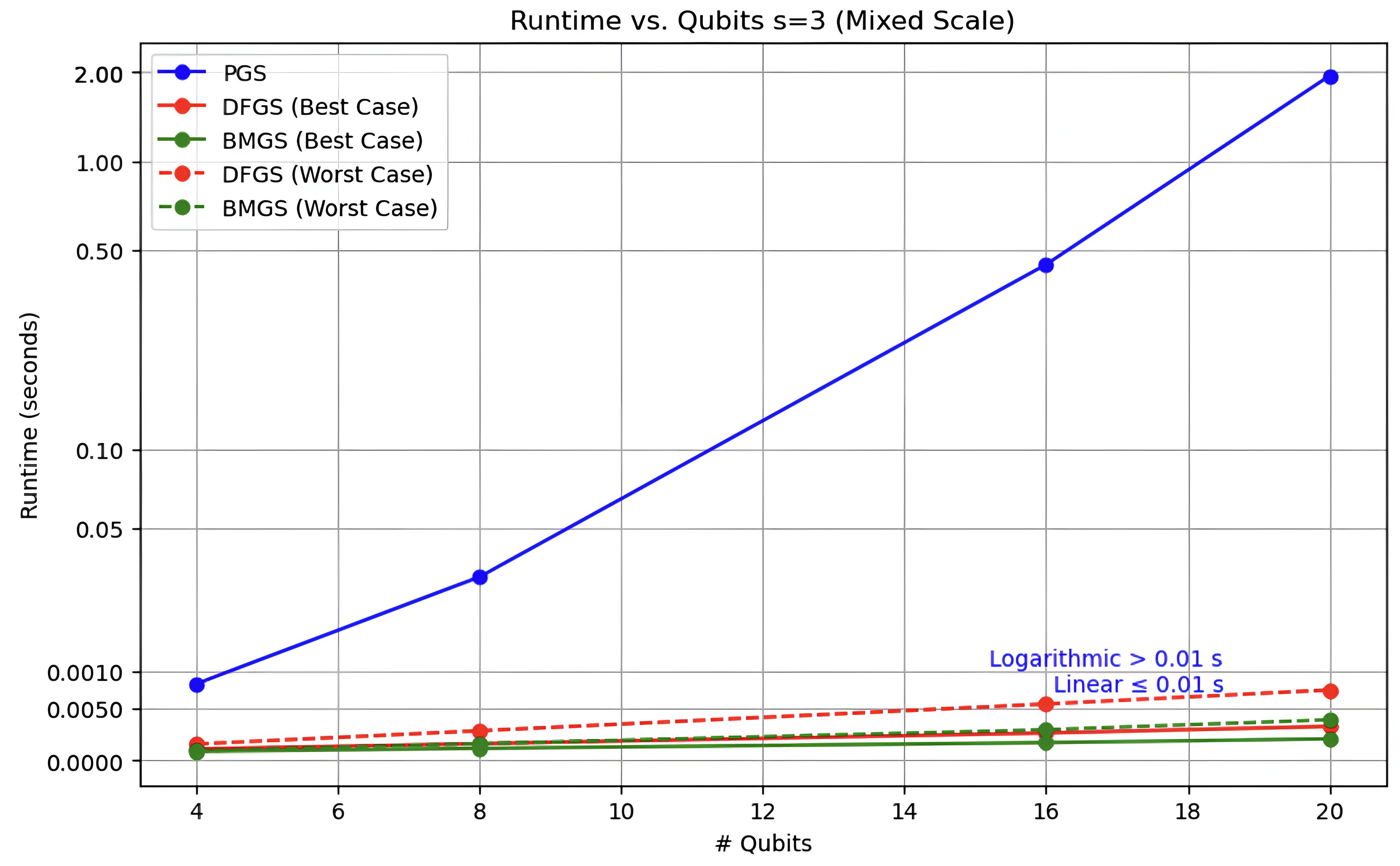}
\caption{Runtime (s) vs. the number of Qubits for (left) $s=1$, (middle) $s=2$, and (right) $s=3$ search space using the proposed BMGS, DFGS~\cite{guo2022}, and PGS~\cite{grover2005} in the best-case scenarios.} 
\label{fig:qubit}
\end{figure*}
Our BMGS procedure performs partial or full Grover searches over intervals of size at most $b$ for each query at level $\lambda$. It is interesting to note that our BMGS is not a collection of isolated Grover searches over smaller subspaces. Instead, the proposed BMGS builds a coordinated, layered search over the entire tree-like space, leveraging PGS across qubit blocks in parallel rather than in full-segment isolation. Hence, the total complexity of the proposed BMGS is evaluated as follows:
\begin{equation}
\begin{split}
\mathcal{I}_{BMGS} = \underbrace{\sqrt{\mathcal{N}}\sum_{\lambda =0}^{\lfloor \frac{r}{dk}\rfloor-1}
\mu_\lambda\left (\sqrt{\frac{1}{b^\lambda}}-\sqrt{\frac{1}{b^{\lambda +1}}}\right)}_{PGS~for~d-segments}\ .
\end{split}
\label{eq:bmgs}
\end{equation}
For each segment of $d$, $\sum_{\lambda =0}^{\lfloor \frac{r}{dk}\rfloor-1}\mathcal{I}_G^\lambda$ number of Oracle calls are required, and the number of Oracle calls is the same for each segment, 
$\sum_{\lambda =0}^{\lfloor \frac{r}{dk}\rfloor-1}\mathcal{I}_G^\lambda$. 
The expected number of PGS performed on layer $\lambda$, denoted by $\mu_\lambda$, plays a crucial role in multi-solution search algorithms. When $\mu_\lambda =1$, indicating a scenario with only one solution, the proposed BMGS algorithm exhibits a complexity of $\mathcal{O}$($\sqrt{\mathcal{N}}$). 
\begin{eqnarray}
\mathcal{I}_{BMGS}^{s=1} &=& \sqrt{\mathcal{N}} \sum_{\lambda =0}^{\lfloor \frac{r}{dk}\rfloor-1}
\mu_\lambda\left (\sqrt{\frac{1}{b^\lambda}}-\sqrt{\frac{1}{b^{\lambda +1}}}\right)\nonumber\\
&=& 
\sqrt{\mathcal{N}}\left (1- \sqrt{\frac{1}{b^{\lfloor \frac{r}{dk}\rfloor}}}\right)
= 
\mathcal{O}(\sqrt{\mathcal{N}})
\label{eq:smgs}
\end{eqnarray}
Conversely, in cases where all elements are solutions ($\mu_\lambda =b^\lambda$), BMGS demonstrates an asymptotic complexity of $\mathcal{O}$($\sqrt{b}\mathcal{N}$). To determine the average complexity, $\mu_\lambda$ should align with the anticipated number of branches containing solutions. This is quantified by defining $\zeta_i$ as follows:
\begin{equation}
\zeta_i = 
\begin{cases}
1 & \text{for branches devoid of solutions} \\
0 & \text{for branches with any number of solutions}
\end{cases}
\end{equation}
here, $i$ corresponds to $i^{th}$ branch on a current layer $\lambda$. The total number of branches without solutions, denoted by $\mathcal{X}_B =\sum_i \zeta_i$.
\begin{equation}
    E(\mathcal{X}_B) = E(\sum_i^{b^\lambda} \zeta_i) = \sum_i^{b^\lambda} E(\zeta_i) = b^\lambda (1-b^{-\lambda})^s\,,
\end{equation}
where $s$ is the arbitrary number of solutions. Since $s$ solutions are distributed across $d$ segments, the probability of a branch being empty should depend on the expected number of solutions:
Hence, $\mu_\lambda = b^\lambda (1-(1-b^{-\lambda})^s$ and Equation~\ref{eq:bmgs} can be rewritten applying Bernoulli inequality $(1 + x)^n \geq 1 + nx$, we have $(1-(1-b^{-\lambda})^s) \leqslant sb^{-\lambda}$ and
\begin{eqnarray}
&\mathcal{I}_{BMGS}^s = 
\sqrt{\mathcal{N}}\sum_{\lambda =0}^{\lfloor \frac{r}{dk}\rfloor-1}b^\lambda (1-(1-b^{-\lambda})^s(\sqrt{\frac{1}{b^\lambda}}\nonumber\\
&-\sqrt{\frac{1}{b^{\lambda +1}}})\leqslant \sqrt{\mathcal{N}}\sum_{\lambda =0}^{\lfloor \frac{r}{dk}\rfloor-1} s(\sqrt{\frac{1}{b^\lambda}}-\sqrt{\frac{1}{b^{\lambda +1}}})\ .
\label{eq:ebmgs}
\end{eqnarray}
To reduce the effective iteration count for multiple solutions, we leverage parallel execution of BMGS over its $d$ segments. Instead of treating all $s$ solutions separately (which leads to the naive bound $\mathcal{O}(s\sqrt{\mathcal{N}})$), we assume solutions are distributed across the search space. Hence, 
\begin{equation}
\begin{split}
\mathcal{I}_{BMGS}^{s} \approx \frac{s}{\sqrt{d}}\sqrt{\mathcal{N}}
= \mathcal{O}(\frac{s}{\sqrt{d}}\sqrt{\mathcal{N}})\ .
   \end{split}
\end{equation}
For $d\geq s$, our BMGS uses at least $d=\mathcal{O}(s)$ segments, then
\begin{equation}
\mathcal{I}_{BMGS}^{s} \approx  \mathcal{O}(\frac{s}{\sqrt{d}}\sqrt{\mathcal{N}}) = \mathcal{O}(\sqrt{s\mathcal{N}})\ .   
\end{equation}
We deduce that the average number of Oracle calls of BMGS for a single solution is $\mathcal{I}_{BMGS}^{s=1}=\sqrt{\mathcal{N}}\left (1- \sqrt{\frac{1}{b^{\lfloor\frac{r}{dk}\rfloor}}}\right)$  and the total time complexity for the multi-solution search is evaluated as $\mathcal{O}$($\sqrt{s\mathcal{N}}$) same as PGS. \textcolor{black}{The multi-solution complexity $\mathcal{O}(\sqrt{s\mathcal{N}})$ is an average-case result under the assumptions that (i) solutions are reasonably distributed across segments, and (ii) segment-wise searches can be executed with effective parallelism. It is not a strict worst-case asymptotic improvement over Grover’s lower bound.}
\begin{theorem}
BMGS complexity is monotonically decreasing in $d$, for $d < r/k$   
 \label{appndx:d_opt}
\end{theorem} 
\textit{Proof}:
\textcolor{black}{We now explicitly state that the monotonic decrease with respect to the number of segments $d$ holds only approximately for $d < \frac{r}{k}$, since floor effects, segment overhead, and auxiliary-qubit costs prevent strict monotonicity in practice.}

\color{black}
\subsection{Query Definition, and Gate Complexity Analysis}
\label{sec:query}

Let the search space consist of $\mathcal{N} = 2^r$ basis states over an $r$-qubit register $\mathcal{Q}$, and let the standard Grover Boolean oracle be defined as
\begin{equation}
 \mathcal{O}_f(x) =
\begin{cases}
1, & x \in \mathcal{M}, \\
0, & x \notin \mathcal{M},
\end{cases}   
\end{equation}
where $\mathcal{M}$ denotes the set of marked states.\\
The corresponding Grover phase oracle is
\begin{equation}
\mathcal{U}_f |x\rangle = (-1)^{\mathcal{O}_f(x)} |x\rangle.
\end{equation}

\subsubsection{Segmented Oracle Construction in BMGS}
In BMGS, the $r$ qubits are divided into $d$ equal segments, each containing
$S_d = \left\lfloor \frac{r}{d} \right\rfloor$ qubits. Let
\begin{equation}
\mathcal{Q} = \bigcup_{i=1}^{d} S_i, \qquad |S_i| = S_d,   
\end{equation}
where $S_i$ denotes the $i$-th segment.\\
For each segment, the local oracle acts only on a window of $k = \lceil \log_2 b \rceil$ qubits corresponding to the branching factor $b$. Thus, the segmented oracle is defined as
\begin{equation}
\mathcal{U}_f^{(i)} |x_{S_i}\rangle=(-1)^{O_f^{(i)}(x_{S_i})}|x_{S_i}\rangle,
\end{equation}
where $\mathcal{O}_f^{(i)}(x_{S_i}) \subseteq O_f(x)$.
The complete BMGS query is therefore
\begin{equation}
\mathcal{Q}_{\text{BMGS}} = \sum_{i=1}^{d}\sum_{\lambda=0}^{\left\lfloor \frac{r}{dk} \right\rfloor -1}
\mathcal{U}_f^{(i,\lambda)},
\end{equation}
which represents the aggregate of all segment-wise oracle invocations required to simulate one logical query to the global oracle.\\

\subsubsection{MCX Gate Complexity: Standard Grover vs BMGS}
In standard Grover Search (GS), the oracle for an $r$-qubit marked state requires an $(r)$-controlled NOT gate (MCX), whose decomposition cost scales as $\mathcal{G}_{\text{MCX}}^{\text{GS}}\approx \mathcal{O}(r)\quad \text{Toffoli-depth}$ with ancilla-assisted decomposition, and $\mathcal{G}_{\text{MCX}}^{\text{GS}}\approx \mathcal{O}(r^2)\quad\text{without ancilla}$.\\
Similarly, the diffuser requires another $r$-controlled operation, giving a total gate cost $\mathcal{G}_{\text{GS}}=2\mathcal{G}_{\text{MCX}}^{\text{GS}} \approx\mathcal{O}(r)$.
Since the number of Grover iterations is $\mathcal{I}_{\text{GS}} =\frac{\pi}{4}\sqrt{\mathcal{N}}$, the total circuit depth becomes $\mathcal{D}_{\text{GS}}
\approx \mathcal{O}\left(r\sqrt{\mathcal{N}}\right)$.\\

\subsubsection{Gate Complexity in BMGS}
In BMGS, each oracle acts only on $k$ qubits, where typically $k = 2
\quad \text{or} \quad k = \lceil \log_2 b \rceil$. Thus, each local oracle requires only a constant-size Toffoli gate: $G_{\text{MCX}}^{\text{BMGS}} \approx \mathcal{O}(k) \approx \mathcal{O}(1)$.\\
The total number of Oracle applications is $\mathcal{I}_{\text{BMGS}}=\sqrt{\mathcal{N}}\left(1 -\sqrt{\frac{1}{b^{\left\lfloor \frac{r}{dk} \right\rfloor}}}\right)$.\\
Hence, the total BMGS circuit depth becomes
\begin{equation}
\mathcal{D}_{\text{BMGS}} \approx \mathcal{O}\left(\mathcal{I}_{\text{BMGS}} \cdot k \right) = \mathcal{O}\left(\sqrt{\mathcal{N}}\left(1 -\sqrt{\frac{1}{b^{\left\lfloor \frac{r}{dk} \right\rfloor}}}\right)\right). 
\end{equation}
Since $k \ll r$ and typically $k=2$, we obtain $D_{\text{BMGS}} \ll \mathcal{D}_{\text{GS}}$.\\
Hence, our BMGS may not improve the worst-case query lower bound
$\mathcal{O}(\sqrt{\mathcal{N}})$, but it significantly reduces MCX control complexity, Toffoli depth, and total circuit depth. This makes BMGS primarily a circuit-efficient and NISQ-friendly reformulation of Grover-style search rather than a strict asymptotic improvement in query complexity.

\color{black}
\section{Simulation Results} 
\label{results:simul}

\subsection{Parameter Selection and Experimental Settings}
\label{exp:sett}

The experimental setup involved conducting multiple trials (maximum $50$) for different solutions ($s=1, 2,$ and $3$) and segments ($d=2$ to $10$) with best and worst-case scenarios using the proposed algorithms BMGS, DFGS~\cite{guo2022}, and PGS~\cite{grover2005} in quantum simulation using the Qiskit Aer simulator. The best-case and worst-case arise when there is the most probable overlap and no overlap among the solutions, respectively.
The simulations are performed on a qubit space ranging from $4$ to $20$ qubits, with each execution consisting of $1024$ shots on $8$-core systems operating at a maximum frequency of $3.5$ GHz and equipped with $8$ GB RAM. In the quantum simulation settings, the Oracle and Diffuser search for $k=2$ qubits are implemented as part of a partial search in the forward and backward directions for each segment within the proposed BMGS algorithm. The choice of $k=2$ qubits has been found to work effectively in a database with an even distribution. Additionally, a DFGS algorithm based on partial search in a depth-wise manner without intercepting qubits or segmentation is implemented using the same quantum simulation settings. \\
\textcolor{black}{To improve clarity and reproducibility, we explicitly specify the parameter settings used in all simulation results. For the comparative runtime and iteration plots in Figure~\ref{fig:qubit} and Tables~\ref{table:BMGS_best-case}-\ref{table:BMGS_worst-case}, we use a fixed segmentation parameter $d = 2$, which corresponds to dividing the $r$-qubit search space into two equal segments for forward and backward Partial Grover Search. This setting provides the most direct comparison with DFGS while preserving the bi-directional structure of BMGS.\\
The branching factor is chosen as $b = 4$, which implies $k = \left\lceil \log_2 b \right\rceil = 2$. The choice of $k = 2$ allows each local oracle and diffuser to operate on only two qubits at a time, enabling implementation using standard Toffoli gates rather than large multi-controlled (MCX) gates. This significantly reduces circuit depth and improves the feasibility of shallow, NISQ-compatible quantum circuits. Empirically, $k = 2$ provides a good balance between search granularity and oracle efficiency for evenly distributed search spaces.\\
In the separate scalability analysis shown in Figure~\ref{fig:RT_vs_d}, the number of segments is varied over $d \in \{2,3,\dots,10\}$, to study the effect of segmentation on runtime and iteration count. Increasing $d$ reduces the effective search depth of each segment since $S_d=\left\lfloor \frac{r}{d}\right\rfloor$, which decreases the number of required Partial Grover layers $L=\left\lfloor\frac{S_d}{k}\right\rfloor$.\\
As a result, increasing $d$ generally reduces runtime and the depth of the oracle for shallow circuits. However, this improvement holds only while $dk < r$, since excessively large $d$ introduces auxiliary-qubit overhead, floor-function effects, and reduced parallel efficiency, which can limit further gains.\\
Thus, the fixed-$d$ experiments evaluate a direct algorithmic comparison, while the variable-$d$ analysis demonstrates the scalability of BMGS as segmentation increases.}\\
\begin{table*}
\scriptsize
\centering
 \caption{Comparative performance analysis of the proposed BMGS, PGS~\cite{grover2005}, and DFGS~\cite{guo2022} for multi-solutions ($s$) with $1024$ shots [Time ($t$) is $\times 10^{-3}$ scale] for the Best-case scenarios.}
\resizebox{\textwidth}{!}{%
\begin{tabular}{p{50pt}p{20pt}p{1pt}p{40pt}p{20pt}p{40pt}p{20pt}p{1pt}p{40pt}p{20pt}p{40pt}p{20pt}p{1pt}p{40pt}p{20pt}p{40pt}p{20pt}}
\hline
\multicolumn{2}{c}{\centering{$\bf{Best-case}$}}  &  & \multicolumn{4}{p{120pt}}{\centering{\textbf{PGS}}} & &\multicolumn{4}{p{120pt}}{\centering{\textbf{DFGS}}} & &\multicolumn{4}{p{120pt}}{\centering{\textbf{BMGS}}} \\
& & & \multicolumn{2}{c}{\centering{$s=2$}} & \multicolumn{2}{p{60pt}}{\centering{$s=3$}} & & \multicolumn{2}{p{60pt}}{\centering{$s=2$}} & \multicolumn{2}{p{60pt}}{\centering{$s=3$}} & & \multicolumn{2}{p{60pt}}{\centering{$s=2$}} & \multicolumn{2}{p{60pt}}{\centering{$s=3$}} \\  
\textbf{\#Qubits ($r$)} & \textbf{\#Trials} & & $Acc.$ & $t(s)$ & $Acc.$ & $t(s)$ & & $Acc.$ & $t(s)$ & $Acc.$ & $t(s)$ & & $Acc.$ & $t(s)$ & $Acc.$ & $t(s)$\\
     \hline
& 10 & & $92.7$ & $4.73$ & $89.8$ & $8.12$ & & $100$ & $0.745$ & $100$ & $0.970$ & & $100$ & $0.520$ & $100$ & $0.739$\\ 
& 20 & & $92.6$ & $5.34$ & $89.9$ & $8.47$ & & $100$ & $0.725$ & $100$ & $0.911$ & & $100$ & $0.543$ & $100$ & $0.740$\\ 
4 & 30 & & $92.5$ & $5.00$ & $90.8$ & $7.77$ & & $100$ & $0.743$ & $100$ & $0.950$ & & $100$ & $0.523$ & $100$ & $0.736$\\
& 40 & & $92.7$ & $5.08$ & $89.7$ & $7.98$ & & $100$ & $0.738$ & $100$ & $0.955$ & & $100$ & $0.510$ & $100$ & $0.734$\\
& 50 & & $92.4$ & $5.21$ & $89.7$ & $7.98$ & & $100$ & $0.736$ & $100$ & $0.905$ & & $100$ & $0.511$ & $100$ & $0.739$\\
& Avg. & & $\bf{92.6\pm0.1}$ & $5.07$ & $\bf{90.0\pm0.4}$ & $8.06$ & & $\bf{100\pm0}$ & $0.737$ & $\bf{100\pm0}$ & $0.938$ & & $\bf{100\pm0}$ & $0.521$ & $\bf{100\pm0}$ & $0.738$\\
\hline
& 10 & & $100$ & $22.5$ & $100$ & $32.8$ & & $100$ & $1.43$ & $100$ & $1.39$ & & $100$ & $0.738$ & $100$ & $0.976$\\ 
& 20 & & $100$ & $22.5$ & $100$ & $30.6$ & & $100$ & $1.15$ & $100$ & $1.40$ & & $100$ & $0.764$ & $100$ & $0.996$\\ 
8 & 30 & & $100$ & $22.5$ & $100$ & $31.3$ & & $100$ & $1.14$ & $100$ & $1.38$ & & $100$ & $0.748$ & $100$ & $0.992$\\
& 40 & & $100$ & $22.6$ & $100$ & $30.4$ & & $100$ & $1.13$ & $100$ & $1.39$ & & $100$ & $0.735$ & $100$ & $0.977$\\
& 50 & & $100$ & $21.5$ & $100$ & $30.8$ & & $100$ & $1.14$ & $100$ & $1.40$ & & $100$ & $0.741$ & $100$ & $0.986$\\
& Avg. & & $\bf{100\pm0}$ & $22.3$ & $\bf{100\pm0}$ & $31.2$ & & $\bf{100\pm0}$ & $1.20$ & $\bf{100\pm0}$ & $1.39$ & & $\bf{100\pm0}$ & $0.745$ & $\bf{100\pm0}$ & $0.986$\\
\hline
& 10 & & $100$ & $33.0$ & $100$ & $43.1$ & & $100$ & $2.30$ & $100$ & $2.40$ & & $100$ & $1.26$ & $100$ & $1.52$\\ 
& 20 & & $100$ & $31.9$ & $100$ & $44.2$ & & $100$ & $2.21$ & $100$ & $2.38$ & & $100$ & $1.23$ & $100$ & $1.46$\\ 
16 & 30 & & $100$ & $31.8$ & $100$ & $45.8$ & & $100$ & $2.33$ & $100$ & $2.38$ & & $100$ & $1.23$ & $100$ & $1.48$\\
& 40 & & $100$ & $31.6$ & $100$ & $45.4$ & & $100$ & $2.15$ & $100$ & $2.34$ & & $100$ & $1.25$ & $100$ & $1.48$\\
& 50 & & $100$ & $31.4$ & $100$ & $45.1$ & & $100$ & $2.25$ & $100$ & $2.35$ & & $100$ & $1.30$ & $100$ & $1.47$\\
& Avg. & & $\bf{100\pm0}$ & $31.9$ & $\bf{100\pm0}$ & $44.7$ & & $\bf{100\pm0}$ & $2.25$ & $\bf{100\pm0}$ & $2.37$ & & $\bf{100\pm0}$ & $1.25$ & $\bf{100\pm0}$ & $1.48$\\
\hline
& 10 & & $100$ & $130$ & $100$ & $195$ & & $100$ & $2.84$ & $100$ & $2.97$ & & $100$ & $1.55$ & $100$ & $1.81$\\ 
& 20 & & $100$ & $128$ & $100$ & $193$ & & $100$ & $2.75$ & $100$ & $3.02$ & & $100$ & $1.58$ & $100$ & $1.83$\\ 
20 & 30 & & $100$ & $129$ & $100$ & $195$ & & $100$ & $2.80$ & $100$ & $3.08$ & & $100$ & $1.56$ & $100$ & $1.81$\\
& 40 & & $100$ & $129$ & $100$ & $194$ & & $100$ & $2.76$ & $100$ & $3.02$ & & $100$ & $1.57$ & $100$ & $1.82$\\
& 50 & & $100$ & $130$ & $100$ & $195$ & & $100$ & $2.77$ & $100$ & $2.98$ & & $100$ & $1.56$ & $100$ & $1.83$\\
& Avg. & & $\bf{100\pm0}$ & $129$ & $\bf{100\pm0}$ & $194$ & & $\bf{100\pm0}$ & $2.78$ & $\bf{100\pm0}$ & $3.01$ & & $\bf{100\pm0}$ & $1.56$ & $\bf{100\pm0}$ & $1.82$\\
\hline
\end{tabular}
}
\label{table:BMGS_best-case}
\end{table*}
The simulation results indicated a significant improvement in runtime without compromising the accuracy of our BMGS search algorithm, compared to the DFGS and PGS algorithms, in both the best and worst cases, as provided in Table~\ref{table:BMGS_best-case} and~Table~\ref{table:BMGS_worst-case}, respectively. The runtime of the PGS algorithm has been observed to grow exponentially with an increasing search space, while the DFGS and BMGS algorithms exhibited linear growth, as shown in Figure~\ref{fig:qubit}. Furthermore, the number of iterations required increased with the number of qubits, with DFGS and BMGS significantly reducing the number of iterations, making them more efficient for larger search spaces, as illustrated in Figure~3 (\emph{Appendix}). 
For a search space of $20$ qubits, the PGS algorithm required $1608$ and $2412$ iterations for $s=2$ and $s=3$, respectively. In comparison, DFGS completed the search in only $20$ iterations for $s=2$ and $30$ iterations for $s=3$ in the worst case, and in the best case in $11$ iterations for $s=2$, and $12$ iterations for $s=3$. Furthermore, BMGS completed the search in $10$ iterations for $s=2$ and $15$ for $s=3$ in the worst case, and in the best case, $6$ for $s=2$ and $7$ for $s=3$.\\
In addition, an increase in segments ($d$) reduces the running time in the worst case, as shown in Figure~4 (\emph{Appendix}). 
The number of iterations escalates exponentially with the qubit count in the search space in standard GS, leading to a proportional increase in the required number of gates. For example, in an $8$-qubit search space, standard GS requires $20$ iterations, while BMGS requires only $2$. Similarly, in a $20$-qubit search space, the standard GS requires $804$ iterations, whereas BMGS requires only $5$, significantly alleviating hardware demands. However, setting $d=10$ in a $20$ qubit search with BMGS reduces the number of iterations to $1$, resulting in a significantly reduced runtime compared to PGS.\\
Standard GS involves applying an oracle and a diffuser to the entire search space, leading to an exponential increase in the number of iterations required as the number of qubits grows. However, our BMGS and DFGS utilize smaller oracles and diffusers on segments of qubits, typically $2$ qubits at a time. This segmented approach results in a linear increase in the number of iterations needed, offering a more efficient search method than the standard GS algorithm. BMGS and DFGS reduce the number of iterations required and demonstrate enhanced parallelism. In BMGS, for instance, oracles and diffusers are applied simultaneously to 2-qubit segments, requiring that each parallel segment include an auxiliary qubit. This parallel processing capability significantly streamlines the search process, ensuring that only one iteration is needed regardless of the search space's size. However, when considering the parallel execution of multiple GS instances ($d$ instances), our BMGS algorithm offers a novel approach in which each segment is processed depth-first, obviating the need for a merging step and enabling direct attainment of the final solution. In the simulation, in a $20$-qubit search space scenario, the standard GS requires around $804$ iterations, DFGS only $10$, and BMGS further reduces this to $5$. This reduction in iterations directly translates into substantially lower circuit depth, showcasing the practical advantages of these refined search methodologies in quantum computing applications.
\begin{table*}
\scriptsize
\centering
 \caption{Comparative performance analysis of the proposed BMGS, PGS~\cite{grover2005}, and DFGS~\cite{guo2022} for multi-solutions ($s$) with $1024$ shots [Time ($t$) is $\times 10^{-3}$ scale] for the Worst-case scenarios.}
\resizebox{\textwidth}{!}{%
\begin{tabular}{p{50pt}p{20pt}p{1pt}p{40pt}p{20pt}p{40pt}p{20pt}p{1pt}p{40pt}p{20pt}p{40pt}p{20pt}p{1pt}p{40pt}p{20pt}p{40pt}p{20pt}}
\hline
\multicolumn{2}{p{70pt}}{\centering{$\bf{Worst-case}$}}  &  & \multicolumn{4}{p{120pt}}{\centering{\textbf{PGS}}} & &\multicolumn{4}{p{120pt}}{\centering{\textbf{DFGS}}} & &\multicolumn{4}{p{120pt}}{\centering{\textbf{BMGS}}} \\
& & & \multicolumn{2}{p{60pt}}{\centering{$s=2$}} & \multicolumn{2}{p{60pt}}{\centering{$s=3$}} & & \multicolumn{2}{p{60pt}}{\centering{$s=2$}} & \multicolumn{2}{p{60pt}}{\centering{$s=3$}} & & \multicolumn{2}{p{60pt}}{\centering{$s=2$}} & \multicolumn{2}{p{60pt}}{\centering{$s=3$}} \\ 
\textbf{\#Qubits($r$)} & \textbf{\#Trials} & & $Acc.$ & $t(s)$ & $Acc.$ & $t(s)$ & & $Acc.$ & $t(s)$ & $Acc.$ & $t(s)$ & & $Acc.$ & $t(s)$ & $Acc.$ & $t(s)$\\
\hline
& 10 & & $92.7$ & $4.73$ & $89.8$ & $8.12$ & & $100$ & $0.913$ & $100$ & $1.38$ & & $100$ & $0.510$ & $100$ & $0.775$\\ 
& 20 & & $92.6$ & $5.34$ & $89.9$ & $8.47$ & & $100$ & $0.910$ & $100$ & $1.32$ & & $100$ & $0.518$ & $100$ & $0.778$\\ 
4 & 30 & & $92.5$ & $5.00$ & $90.8$ & $7.77$ & & $100$ & $0.947$ & $100$ & $1.39$ & & $100$ & $0.517$ & $100$ & $0.753$\\
& 40 & & $92.7$ & $5.08$ & $89.7$ & $7.98$ & & $100$ & $0.933$ & $100$ & $1.36$ & & $100$ & $0.518$ & $100$ & $0.773$\\
& 50 & & $92.4$ & $5.21$ & $89.7$ & $7.98$ & & $100$ & $0.918$ & $100$ & $1.34$ & & $100$ & $0.520$ & $100$ & $0.770$\\
& Avg. & & $\bf{92.6\pm0.1}$ & $5.07$ & $\bf{90.0\pm0.4}$ & $8.06$ & & $\bf{100\pm0}$ & $0.924$ & $\bf{100\pm0}$ & $1.36$ & & $\bf{100\pm0}$ & $0.517$ & $\bf{100\pm0}$ & $0.770$\\
\hline
& 10 & & $100$ & $22.5$ & $100$ & $32.8$ & & $100$ & $1.70$ & $100$ & $2.60$ & & $100$ & $0.943$ & $100$ & $1.38$\\ 
& 20 & & $100$ & $22.5$ & $100$ & $30.6$ & & $100$ & $17.3$ & $100$ & $2.58$ & & $100$ & $0.947$ & $100$ & $1.37$\\ 
8 & 30 & & $100$ & $22.5$ & $100$ & $31.3$ & & $100$ & $1.72$ & $100$ & $2.59$ & & $100$ & $0.935$ & $100$ & $1.39$\\
& 40 & & $100$ & $22.6$ & $100$ & $30.4$ & & $100$ & $1.73$ & $100$ & $2.58$ & & $100$ & $0.941$ & $100$ & $1.39$\\
& 50 & & $100$ & $21.5$ & $100$ & $30.8$ & & $100$ & $1.73$ & $100$ & $2.59$ & & $100$ & $0.932$ & $100$ & $1.39$\\
& Avg. & & $\bf{100\pm0}$ & $22.3$ & $\bf{100\pm0}$ & $31.2$ & & $\bf{100\pm0}$ & $1.72$ & $\bf{100\pm0}$ & $2.59$ & & $\bf{100\pm0}$ & $0.940$ & $\bf{100\pm0}$ & $1.38$\\
\hline
& 10 & & $100$ & $33.0$ & $100$ & $43.1$ & & $100$ & $36.3$ & $100$ & $5.65$ & & $100$ & $1.86$ & $100$ & $2.73$\\ 
& 20 & & $100$ & $31.9$ & $100$ & $44.2$ & & $100$ & $36.3$ & $100$ & $5.41$ & & $100$ & $1.85$ & $100$ & $2.65$\\ 
16 & 30 & & $100$ & $31.8$ & $100$ & $45.8$ & & $100$ & $3.71$ & $100$ & $5.47$ & & $100$ & $1.84$ & $100$ & $2.69$\\
& 40 & & $100$ & $31.6$ & $100$ & $45.4$ & & $100$ & $3.60$ & $100$ & $5.36$ & & $100$ & $1.85$ & $100$ & $2.68$\\
& 50 & & $100$ & $31.4$ & $100$ & $45.1$ & & $100$ & $3.63$ & $100$ & $5.43$ & & $100$ & $1.85$ & $100$ & $2.69$\\
& Avg. & & $\bf{100\pm0}$ & $31.9$ & $\bf{100\pm0}$ & $44.7$ & & $\bf{100\pm0}$ & $3.64$ & $\bf{100\pm0}$ & $5.47$ & & $\bf{100\pm0}$ & $1.85$ & $\bf{100\pm0}$ & $2.69$\\
\hline
& 10 & & $100$ & $130$ & $100$ & $195$ & & $100$ & $4.80$ & $100$ & $7.39$ & & $100$ & $2.49$ & $100$ & $3.70$\\ 
& 20 & & $100$ & $128$ & $100$ & $193$ & & $100$ & $4.92$ & $100$ & $7.20$ & & $100$ & $2.51$ & $100$ & $3.67$\\ 
20 & 30 & & $100$ & $129$ & $100$ & $195$ & & $100$ & $4.78$ & $100$ & $7.19$ & & $100$ & $2.48$ & $100$ & $3.70$\\
& 40 & & $100$ & $129$ & $100$ & $194$ & & $100$ & $5.13$ & $100$ & $7.16$ & & $100$ & $2.50$ & $100$ & $3.69$\\
& 50 & & $100$ & $130$ & $100$ & $195$ & & $100$ & $4.79$ & $100$ & $7.21$ & & $100$ & $2.50$ & $100$ & $3.70$\\
& Avg. & & $\bf{100\pm0}$ & $129$ & $\bf{100\pm0}$ & $194$ & & $\bf{100\pm0}$ & $4.89$ & $\bf{100\pm0}$ & $7.23$ & & $\bf{100\pm0}$ & $2.50$ & $\bf{100\pm0}$ & $3.69$\\
\hline
\end{tabular}
}
\label{table:BMGS_worst-case}
\end{table*}
\color{black}
\begin{table}[h]
\centering
\caption{Comparison of circuit-level complexity for standard Grover Search (GS), Depth-First Grover Search (DFGS), and the proposed BMGS for a 20-qubit search space ($r=20$), with branching factor $b=4$, $k=2$, and $d=2$. Here, $\mathcal{I}$ denotes the number of Grover iterations, $\mathcal{D}$ denotes approximate total circuit depth, and $\mathcal{T}$-depth denotes the effective Toffoli-depth after MCX decomposition.}
\begin{tabular}{|c|c|c|c|c|}
\hline
\bf{Algorithm} & \bf{Oracle Size} & \bf{Iterations} $(\bf{\mathcal{I}})$ & \bf{Circuit Depth} $(\bf{\mathcal{D}})$ & \bf{Approx. $\mathcal{T}$-depth} \\
\hline
GS & $20$-controlled MCX & $804$ & $\sim 16080$ & Very High $(\sim O(20\times804))$ \\
\hline
PGS & $4$-qubit local oracle & $1608$ & $\sim 6432$ & Moderate \\
\hline
DFGS & $2$-qubit segmented oracle & $10$--$20$ & $\sim 40$--$80$ & Low \\
\hline
BMGS & $2$-qubit segmented oracle & $5$--$10$ & $\sim 20$--$40$ & Lowest \\
\hline
\end{tabular}
\label{tab:depth_comparison}
\end{table}
\\ We also present numerical comparisons of circuit complexity metrics in Table~\ref{tab:depth_comparison} to better demonstrate the practical hardware advantage of BMGS beyond simulation runtime alone. For standard Grover Search (GS), the oracle and diffuser require large multi-controlled gates acting on all $r = 20$ qubits, resulting in high circuit depth and large $T$-depth due to MCX decomposition. With approximately $\mathcal{I}_{\text{GS}}=\frac{\pi}{4}\sqrt{2^{20}}\approx 804$ iterations, the total circuit depth becomes prohibitively large for shallow quantum hardware.\\
In contrast, BMGS applies Partial Grover Search on local windows of only $k = 2$ qubits using standard Toffoli gates. For the same search space, BMGS requires only $5\text{--}10$ effective iterations depending on the number of solutions and segment overlap, reducing the circuit depth by several orders of magnitude.\\
This demonstrates that the primary advantage of BMGS lies not in improving Grover’s asymptotic lower bound, but in significantly reducing oracle depth, $\mathcal{T}$-depth, and overall circuit complexity, making the algorithm substantially more suitable for NISQ implementations.

\color{black}
\section{Discussion}
\label{discuss}

In this work, we investigate a novel structured bidirectional multisolution search challenge in which the satisfaction criterion $\mathcal{C} =\prod_{i=1}^{\frac{d}{2}}\mathcal{F}_i \wedge \mathcal{R}_i$ can be decomposed into $\mathcal{F}$ and $\mathcal{R}$ for $d$ segments (2 segments each for $\mathcal{F}$ and $\mathcal{R}$), representing search-based satisfaction conditions. We introduce a scalable and faster quantum partial search method for multiple target items, utilizing smaller oracles on intercepted qubits (segments). The fundamental concept of the proposed BMGS algorithm is to amplify the amplitude of items in each segment that meet $\mathcal{F}$ and $\mathcal{R}$. The maximum number of Grover iterations (Oracles calls) for our BMGS for each solution or marked state evaluated as $\sqrt{\mathcal{N}}\left (1- \sqrt{\frac{1}{b^{\lfloor\frac{r}{dk}\rfloor}}}\right)$ over PGS~\cite{grover2005}, which take $\frac{\pi}{4}\sqrt{\mathcal{N}}\sqrt{1-\frac{1}{b}}$, and DFGS~\cite{guo2022}$\sqrt{\mathcal{N}}\left(1-\sqrt{\frac{1}{b^r}}\right)$. We also show that our BMGS requires fewer iterations (see \emph{Appendix}) and achieves optimal $\mathcal{O}$($\sqrt{s\mathcal{N}}$) average complexity with a higher success probability, as analyzed in \emph{Appendix}. Our simulation results reveal that the proposed BMGS algorithms significantly outperformed DFGS and PGS regarding runtime while maintaining comparable measurement accuracy. A comparison with a number of iterations, efficiency of our BMGS, DFGS, and PGS algorithms is provided in Table~3 (\emph{Appendix}). 
This efficiency is primarily attributed to the effective utilization of gates and Grover iterations in DFGS and BMGS. Moreover, unlike standard GS, where the use of MCX gates becomes increasingly impractical in higher search spaces due to the exponential growth in the number of control qubits, the oracles in DFGS and BMGS remain static in size. For instance, in a 20-qubit search space, the standard GS requires an oracle with an MCX gate applied on $20$ control qubits and $1$ target bit. In contrast, BMGS employs a series of oracles using MCX (Toffoli) gates on $2$ control qubits and $1$ target qubit, resulting in a more resource-efficient and practical algorithm that overcomes the limitations of the standard GS. Employing smaller Oracles on segmented qubits, facilitating the design of Oracles for higher qubits, and ensuring error-free quantum circuit operation, and Toffoli gates in place of MCX gates, DFGS and BMGS offer a more efficient and scalable solution for quantum search algorithms, demonstrating the adaptability and innovation required to address the challenges posed by larger search spaces in quantum computing. The combination of our BMGS algorithm's ability to significantly reduce gate complexity and runtime, and its use of more practical gates, makes quantum search algorithms much more practical for near-term quantum hardware. \\
However, despite the non-requirement of the number of solutions prior in the proposed BMGS algorithm, asymptotically, it attains an optimal search technique ($\sqrt{s\mathcal{N}}$) due to the structured search problems requiring qubits' interceptions into segments and the decomposition of the search space. 

\color{black}
\subsection{Future Extension: BMGS for Grover Adaptive Search}
\label{futur:dirc}

A promising future research direction is the integration of the proposed BMGS framework with Grover Adaptive Search (GAS), particularly for solving large-scale optimization problems formulated as Quadratic Unconstrained Binary Optimization (QUBO) or Ising models.\\
In GAS, Grover’s search is repeatedly applied using a threshold-based oracle that marks candidate states satisfying $f(x) < \theta$, where $f(x)$ is the objective function and $\theta$ is the current best-known threshold. The algorithm iteratively updates $\theta$ as better solutions are found, transforming optimization into a sequence of adaptive Grover searches.\\
BMGS can naturally serve as a circuit-efficient replacement for the standard Grover subroutine in this framework. Instead of applying a global oracle over all variables, the threshold oracle can be decomposed into segmented local oracle evaluations over subsets of qubits, enabling Partial Grover Search with forward and backward exploration across the solution space.\\
This approach is particularly attractive for large QUBO/Ising instances, where oracle construction and large MCX gates become the dominant bottleneck. By replacing large, globally controlled operations with smaller, segmented Toffoli-based oracles, BMGS can significantly reduce circuit depth, $\mathcal{T}$-depth, and hardware overhead, making adaptive Grover optimization more suitable for shallow, NISQ-compatible quantum devices.\\
Therefore, BMGS may function as a practical search layer within adaptive Grover-based optimization frameworks, offering improved implementability while preserving the amplitude amplification advantages of Grover-style optimization. A full theoretical and experimental investigation of BMGS within GAS remains an important direction for future work.

\color{black}
\section{Conclusions}
\label{conclud}

The study presents a novel approach to developing Bidirectional implementations of the partial Grover's quantum search method for any number of solutions or targets, reducing the computational complexity of the number of Oracle calls. Our BMGS algorithm achieves asymptotic performance $\mathcal{O}(\sqrt{\mathcal{N}})$ for single-solution search and can outperform both DFGS and PGS in iteration count while requiring shallower circuits than DFGS. It thus offers a favorable trade-off between speed and hardware feasibility, offering scalable solutions for any qubit count. The framework offers a potential solution for achieving a quantum advantage in implementing the proposed BMGS algorithm without compromising the probability of success in searching multiple targets. The proposed approach achieves efficiency comparable to PGS and DFGS, with fewer iterations and runtime, by utilizing amplitude amplification on smaller oracles within each segment. Consequently, the proposed BMGS algorithm optimizes gate utilization, significantly reducing run times. It is worth noting that the time complexity of our BMGS algorithms reaches optimality (asymptotically), as they exhibit multi-solution time complexity of $\mathcal{O}$($\sqrt{s\mathcal{N}}$) and the number of iterations required in BMGS is notably reduced compared to DFGS and PGS, showcasing the efficiency and effectiveness in quantum search algorithms.

\bibliography{IEEEabrv,Deb}

\appendix

\section{Grover Search} 
\label{grover_search:algorithm}

Grover's algorithm for quantum searching is primarily concerned with searching an unstructured data array of $\mathcal{N}$ items~\cite{grover1996}. Grover's quantum search approach has to locate an element in the collection $\{L=L_1, L_2, L_3,\ldots L_{\mathcal{N}}\}$, where $\mathcal{N}$ is the total number of elements in the set $L$, for it to be effective. Finding an element in the collection is necessary for $\delta(x) \rightarrow[0, 1]$~\cite{mahmud2022}. This is the responsibility of the Boolean function $\delta$. The full Grover search algorithm locates the marked state in $\mathcal{I}_f$ iterations~\cite{grover2005}. 
\begin{equation}
\mathcal{I}_f = \frac{\pi}{4}\sqrt{\mathcal{N}}, \mathcal{N}\longrightarrow \infty\ .
\end{equation}
The input qubits are first converted into a superposition with equal coefficients using a Hadamard gate. The initial quantum state is subjected to phase inversion and inversion around the mean~\cite{mahmud2022}. \\
Grover's iteration, also known as Grover's operator, is used in the quantum Grover search circuit. Grover's iteration is divided into two phases~\cite{grover1996}. In a marked state, the Oracle function first flips the phase of a single amplitude. The diffusion layer is said to have done its job when the indicated state amplitude flips over. The target state is inverted, causing its amplitude to grow dramatically, while the amplitudes of the other states only slightly decrease. The other states maintain their previous amplitude. 

\subsection{Oracle}
An Oracle~\cite{williams2011} is often shown as a black box that inverts the chosen coefficient of the target base state after receiving the input state, $|y\rangle$.\\
\textit{Initialization}: 
Due to creating an equal superposition of base states, the Hadamard gate ($H$ gate) is utilized in the initial stage of Grover's search algorithm to put all of the qubits in superposition~\cite{williams2011}. The $|0\rangle$ state, \emph{i.e.},$\frac{1}{\sqrt{2}}(|0\rangle+|1\rangle)$, is superposed with equal probability over the $|0\rangle$ and $|1\rangle$ states when a $H$ gate is applied to it. The following matrix represents an $H$ gate:
\begin{equation}
    H = \frac{1}{\sqrt{2}} \begin{bmatrix}1 & \phantom{-}1\\ 1 & -1 \end{bmatrix}\ .
\end{equation}
The uniform superposition of every basis vector in the whole database serves as the starting point for the Grover search as follows~\cite{nielsen2001}:
\begin{equation}
    |\mathcal{S}\rangle = \frac{1}{\sqrt{\mathcal{N}}}\sum_{y=0}^{\mathcal{N}-1}|y\rangle, \langle \mathcal{S}|\mathcal{S}\rangle =1\ . 
\end{equation}
Iteratively, Grover's algorithm looks for a single target object $|x\rangle$. A unitary transform is the Grover iteration:
\begin{equation}
    \mathcal{I}_G= -\mathcal{I}_s\mathcal{I}_x\,, 
\end{equation}
Two inversions on the marked state $|x\rangle$ are presented as $\mathcal{I}_s$ (Diffusion operator) and $\mathcal{I}_x$ (Oracle operator) as follows:
\begin{equation}
\mathcal{I}_s = I-2|\mathcal{S}\rangle\langle\mathcal{S}|\,,
\end{equation}
\begin{equation}
\mathcal{I}_x = I-2|x\rangle\langle x|\,,
\end{equation}
where $I$ is an identity operator.\\
The marked state undergoes a phase flip using the Oracle function. All states, including the one indicated, have an initial amplitude of $\frac{1}{\sqrt{\mathcal{N}}}$. All states have a mean amplitude of $\frac{1}{\sqrt{\mathcal{N}}}$. The initial state $\alpha^{(0)}$ has its amplitude transformed by the phase flip to $-\frac{1}{\sqrt{\mathcal{N}}}$. The change in mean amplitude ($\mu$) is defined as follows~\cite{konar2022}:
\begin{equation} 
\begin{split}
\mu^{(0)} = \frac{(\mathcal{N}-1)\beta^{(0)} + \alpha^{(0)}}{\mathcal{N}} =  \frac{(\mathcal{N}-1)\frac{1}{\sqrt{\mathcal{N}}}-\frac{1}{\sqrt{\mathcal{N}}}}{\mathcal{N}} =\frac{\mathcal{N}-2}{\mathcal{N}\sqrt{\mathcal{N}}}\,, 
\end{split}
\label{eq:4}
\end{equation} 
where the amplitudes of the marked and unmarked states are denoted by $\alpha$ and $\beta$, respectively. \\
The Grover iteration $I_G$ rotates at an angle $\omega$ in the Hilbert space, moving from $\mathcal{S}$ to the destination $|x\rangle$, where $\sin^2 \omega = \frac{1}{\mathcal{N}}$ and $t$ iterations are as follows:
\begin{equation}
\mathcal{I}_G^t|\mathcal{S}\rangle = \sin ((2t +1)\omega)|x\rangle + \frac{\cos ((2t +1)\omega)}{\sqrt{\mathcal{N}-1}}\sum_{\mathcal{S}\neq x}^{\mathcal{N}-1}|\mathcal{S}\rangle\ .
\label{eq:I_g}
\end{equation}
Hence, following $\mathcal{I}_f = \frac{\pi}{4\omega}-\frac{1}{2}$ repetitions, the amplitudes of all other items vanish, and the probability amplitude of $|x\rangle$ becomes unity. Hence, 
\begin{equation}
\mathcal{I}_G^t|\mathcal{S}\rangle = |x\rangle\ .
\end{equation}

\subsection{Diffusion}
A diffusion model can be implemented as \emph{HX + Oracle + XH}, where $H$ represents the Hadamard transform and $X$ represents the $Pauli-X$ gate~\cite{konar2022}. This combination inverts the amplitudes surrounding their mean. According to Grover's search, the amplitude of the marked state will increase by $\frac{1}{\sqrt{\mathcal{N}}}$. The marked state $\alpha^{(1)}$ has a new amplitude that can be calculated using formula 2.$\alpha^{(0)}$-$\mu^{(0)}$. We can derive $\alpha^{(1)}$ as follows using $\mu^{(0)}$= $\frac{\mathcal{N}-2}{\mathcal{N}\sqrt{\mathcal{N}}}$ and $\alpha^{(0)}$= -$\frac{1}{\sqrt{\mathcal{N}}}$ (from Equation~\ref{eq:4}):
\begin{equation}
\begin{split}
 \alpha^{(1)}=2\frac{(\mathcal{N}-2)}{\mathcal{N}\sqrt{\mathcal{N}}} +\frac{1}{\sqrt{\mathcal{N}}}= \frac{3(\mathcal{N}-1)}{N\sqrt{\mathcal{N}}}\ .
 \end{split}
\label{eq:6}   
\end{equation} 
\section{GRK's Partial Grover Search Algorithm}
\label{grov:partial}

Grover and Radhakrishnan (GRK)~\cite{grover2005} proposed a faster quantum search method for partial search using the same Oracle as the Grover algorithm. The partial search also begins with the uniform superposition of all base states. The generic partial Grover search problem considers the following: A database with $\mathcal{N}$ items is divided into $b$ identically sized blocks.
\begin{equation}
 \mathcal{B} = \frac{\mathcal{N}}{b}\ .
\end{equation}
In partial Grover search, approximately $\frac{\pi}{4}(1-c(b))\sqrt{\mathcal{N}}$ searches are required to locate the target block~\cite{grover2005}. When the blocks are infinitely large ($\mathcal{B} \longrightarrow \infty$), $c(b)$ is a finite positive number that depends on $b$.\\
Every block has concurrent local Grover iterations, $j$, as follows.
\begin{equation}
 \mathcal{I}_L = \bigoplus_{j=1}^{b}\mathcal{I}_L^j =-(\bigoplus_{j=1}^{b}\mathcal{I}_s^l)\mathcal{I}_x\,,
\end{equation}
where each $\mathcal{I}_L^j = -\mathcal{I}_s^l\mathcal{I}_x$ and inversion on the marked state (query to the oracle) $|x\rangle$ is $\mathcal{I}_x$ and the local inversion is presented as $\mathcal{I}_s^l$ (Diffusion operator)
\begin{equation}
\mathcal{I}_s^l = I-2|\mathcal{S}_l\rangle\langle\mathcal{S}_l|\ .
\end{equation}
The uniform superposition of items in a block, $|\mathcal{S}_l\rangle$, is represented here as follows.
\begin{equation}
    |\mathcal{S}_l\rangle = \frac{1}{\sqrt{\mathcal{B}}}\sum_{i=0}^{\mathcal{B}-1}|z\rangle\ .
\end{equation}
Regional repetition $\mathcal{I}_L$ is the Grover iteration carried out concurrently in every block. Only the target block has a non-trivial operation [rotation] with a new rotation angle $\omega_l$ specified by
\begin{equation}
    \sin^2 \omega_l = \frac{b}{\mathcal{N}} = \frac{1}{\mathcal{B}}\ .
\end{equation}
Every item in non-target blocks can have its amplitudes eliminated by applying one more global iteration~\cite{liao2006}.
\begin{equation}
 \mathcal{D}=\mathcal{I}_G \mathcal{I}_L^j \mathcal{I}_G^t|\mathcal{S}\rangle = \sin \omega|x\rangle + \frac{\cos \omega}{\sqrt{\mathcal{B}-1}}\sum_{\mathcal{S}\neq x}^{\mathcal{B}-1}|\mathcal{S}\rangle\ .
\end{equation}
For GRK partial Grover's search, we randomly select ($b - 1$) blocks and apply Grover's quantum search method to $\mathcal{N}(1- \frac{1}{b})$ places inside the selected blocks. This would require $\sqrt{\mathcal{B}(b-1)} = \frac{\pi}{4}\sqrt{\mathcal{N}}\sqrt{\frac{b-1}{b}}$ inquiries, and it is faster than a standard full Grover search~\cite{grover2005}.

\section{Algorithm 1: Bi-directional Multi-Solution Grover Search (BMGS)}
\label{appndx:algo}

\textbf{Input Parameters}
\begin{itemize}
    \item $k$: Number of bits per search window (Grover block size).
    \item $b$: Branching factor for bi-directional search.
    \item $r$: Total number of qubits (database size $\mathcal{N} = 2^r$).
    \item $x$: List of target elements to search.
    \item $s$: Total number of solutions.
    \item $d$: Number of segments dividing the $r$-qubit register.
    \item $A$: Auxiliary qubits used for tracking found bits ($|A| = d$).
\end{itemize}

\vspace{0.5em}
\textbf{Step 1: Initialization}
\begin{enumerate}
    \item \textbf{Quantum Register Initialization:}
    \begin{itemize}
        \item Prepare a quantum register $Q$ with $r$ qubits initialized to $|0\rangle^r$.
        \item Define the total search space size $\mathcal{N} = 2^r$.
        \item Apply Hadamard gates to all qubits to generate superposition:
        \[
            H^{\otimes r} |0^r\rangle
        \]
    \end{itemize}
    
    \item \textbf{Auxiliary Qubit Initialization:}
    \begin{itemize}
        \item Set $k = \lceil \log_2 b \rceil$, defining how many bits to process per Grover iteration.
        \item Initialize $d$ auxiliary qubits in the state $|0\rangle^d$.
        \item Define $\mathcal{S}_d = \lfloor \frac{r}{d} \rfloor$ to partition the $r$-qubit space into $d$ segments.
        \item Set $L = \lfloor \mathcal{S}_d / k \rfloor$, denoting the number of recursive PGS iterations per segment.
    \end{itemize}
\end{enumerate}

\vspace{0.5em}
\textbf{Step 2: Parallel Segment Processing}
\begin{itemize}
    \item For each segment index $\ell = 0$ to $d - 1$, execute the Grover steps \textbf{in parallel}.
    \item For each layer $\lambda = 0$ to $L - 1$, search for the next $k$ bits within the segment.
    \item Alternate directions: forward search for even $\ell$ (except the last), backward search for odd $\ell$ and always for $\ell = d - 1$.
\end{itemize}

\vspace{0.5em}
\textbf{Step 3: Forward Search Phase}
\begin{itemize}
    \item Executed when $\ell \bmod 2 = 0$ and $\ell \neq d - 1$.
    \item Iterate through qubit indices in the segment from $\ell \cdot \mathcal{S}_d$ to $(\ell + 1) \cdot \mathcal{S}_d - k$, incrementing by $k$.
    \item Each iteration defines a $k$-qubit window: $q = [j, j + k - 1]$.
    \item If $q$ contains one qubit, apply HSQ. Otherwise, apply PGS if the current bit pattern differs from the previous layer.
\end{itemize}

\vspace{0.5em}
\textbf{Step 4: Backward Search Phase}
\begin{itemize}
    \item Executed when $\ell \bmod 2 = 1$ or $\ell = d - 1$.
    \item Iterate from the end of the segment backward: $j = (\ell + 1) \cdot \mathcal{S}_d - 1$ down to $\ell \cdot \mathcal{S}_d$, decrementing by $k$.
    \item Define the $k$-qubit window as $q = [\max(j - k + 1, \ell \cdot \mathcal{S}_d), j]$.
    \item If $q$ contains one qubit, apply HSQ. Otherwise, apply PGS if the window changed from the previous layer.
\end{itemize}

\vspace{0.5em}
\textbf{Step 5: Measurement}
\begin{itemize}
    \item After all segment passes and Grover iterations are complete, measure the quantum register $Q$.
    \item The resulting classical bitstring corresponds to one of the discovered solutions.
\end{itemize}

\textbf{Algorithm 2: HSQ (Handle Single Qubit)}\\
Applies a Hadamard-Phase-Hadamard (H-Z-H) transformation for a single qubit.

\textbf{Steps:}
\begin{enumerate}
    \item Apply a Hadamard gate $H$ to qubit $q[0]$.
    \item If the target bit is '1', apply a $Z$ gate.
    \item Apply another Hadamard gate.
\end{enumerate}

\title{Algorithm 3: PGS (Partial Grover Search)}
\textbf{Purpose:} Performs a Grover search on a selected segment of $k$ qubits at a time.

\textbf{Steps:}
\begin{enumerate}
    \item Apply the oracle to mark the states corresponding to the target element.
    \item Apply the diffusion operator to amplify the marked states.
\end{enumerate}

The BMGS algorithm enhances Grover’s search by combining segmentation and bidirectional search, reducing redundant searches and improving efficiency for large databases.

\color{black}
\subsection{Illustrative Example of BMGS}

Consider a database of size $\mathcal{N} = 2^8 = 256$ represented using $r = 8$ qubits, and assume the marked state is $|x\rangle = |01100110\rangle$.
Let the branching factor be $b = 4$, $\qquad k = \lceil \log_2 b \rceil = 2$,
and divide the search space into $d = 2$ equal segments. Thus, each segment contains $S_d =\left\lfloor \frac{r}{d}\right\rfloor=4$ qubits.\\
The marked state is segmented as $|x\rangle =|0110\rangle \;|\;|0110\rangle$,
where the first segment is searched in the forward direction and the second segment is searched in the backward direction.

\subsubsection{Step 1: First Partial Grover Search}

For the first segment, BMGS applies Partial Grover Search to determine the first $k = 2$ bits: $|00\rangle \rightarrow |01\rangle$.\\
Simultaneously, for the second segment, backward Partial Grover Search determines the last two bits:$|00\rangle\leftarrow|10\rangle$.\\
Thus, after the first layer, the known partial state becomes $|01 \; ?? \; ?? \; 10\rangle$.

\subsubsection{Step 2: Second Partial Grover Search}

The next forward search identifies the remaining two bits of the first segment: $|01\rangle \rightarrow|0110\rangle$.\\
Similarly, the backward search identifies the remaining two bits of the second segment: $|10\rangle \leftarrow |0110\rangle$.\\
Now the full marked state is reconstructed: $|01100110\rangle$.

\subsubsection{Complexity Reduction}

Standard Grover Search on 8 qubits requires approximately $\mathcal{I}_{\text{GS}}=\frac{\pi}{4}\sqrt{256}\approx 12$ iterations using an 8-controlled oracle. In contrast, BMGS performs search using only 2-qubit local oracles (Toffoli gates), requiring approximately $\mathcal{I}_{\text{BMGS}}= \sqrt{256}\left(1-\sqrt{\frac{1}{4^{\left\lfloor
\frac{8}{2\times2}\right\rfloor}}}\right)=16\left(1-\frac{1}{4}\right)=12$
effective local iterations, but with significantly smaller oracle complexity and much shallower circuit depth.\\
This example illustrates how BMGS reduces practical implementation costs by replacing large, multi-controlled operations with smaller, segmented Partial Grover Searches while preserving Grover-style amplitude amplification.

\color{black}

\section{Analysis on Error Probability of BMGS Algorithm}
\label{appndx:error}

The error probability in quantum search algorithms like Grover’s Search (GS), Partial Grover Search (PGS), and Bi-directional Multi-Solution Grover Search (BMGS) arises due to imperfect amplitude amplification and measurement uncertainty. In this section, we mathematically analyze the error probability of BMGS and how it compares to standard GS, PGS algorithms.

\subsection{Error Probability in Grover’s Search}
After $t$ iterations, the probability of measuring the correct marked state is:
\begin{equation}
P_{\text{Grover}}(t) = \sin^2 \left( (2t+1) \theta \right), 
\end{equation}
where:
\begin{equation}
\theta = \arcsin \left(\frac{1}{\sqrt{\mathcal{N}}}\right).    
\end{equation}
The optimal number of iterations $t^*$ is chosen such that $P_{\text{Grover}}(t^*) \approx 1$. If $t$ is off by even 1 iteration, the probability decreases.\\
The error probability is:
\begin{equation}
 \varepsilon_{\text{Grover}} = 1 - P_{\text{Grover}}(t^*).   
\end{equation}

\subsection{Error Probability in Partial Grover Search}
For PGS, the search is performed within a subset of the full database. The probability of success depends on the effective search space size $\mathcal{B}$ (number of blocks):
\begin{equation}
P_{\text{PGS}}(t) = \sin^2 \left( (2t+1) \omega_\mathcal{B} \right),    
\end{equation}
where:
\begin{equation}
 \omega_\mathcal{B} = \arcsin \left(\frac{1}{\sqrt{\mathcal{B}}}\right),   
\end{equation}
and $\mathcal{B} = \frac{\mathcal{N}}{b}$, with $b$ being the branching factor.\\
The error probability is:
\begin{equation}
\varepsilon_{\text{PGS}} = 1 - \sin^2 \left( (2t+1) \arcsin \left(\frac{1}{\sqrt{\mathcal{B}}}\right) \right).  
\end{equation}

\subsection{Error Probability in BMGS}
Thus, the probability of successfully locating a solution in of BMGS is:
\begin{equation}
 P_{\text{BMGS}}(t) = \sin^2 \left( (2t+1) \omega \right),   
\end{equation}
where:
\begin{equation}
\omega = \arcsin \left(\frac{1}{\sqrt{\mathcal{N}}}\right),    
\end{equation}
and $\mathcal{N} = 2^r$ is the number of elements in the full search space.\\
Using the bi-directional approach, the probability of successfully finding a solution across both directions is:
\begin{equation}
P_{\text{BMGS, total}}(t) = 1 - \prod_{i=1}^{d} (1 - P_{\text{BMGS}, i}(t)). 
\end{equation}
\textcolor{black}{In the proposed BMGS algorithm, the search space is processed using segmented oracle operations over multiple qubit blocks. However, these segment-wise operations do not imply probabilistic independence between segments, since the amplitudes remain globally coupled in the full Hilbert space. The quantum state is shared across the entire computational basis, and amplitude amplification occurs globally rather than independently within each segment.\\
Therefore, the expression $P_{\text{BMGS, total}}(t)$ is used only as a heuristic approximation for empirical estimation when inter-segment amplitude coupling is weak and shallow segmented circuits are considered. Accordingly, we emphasize that the rigorous success probability of our BMGS remains governed by the standard global Grover evolution. Thus, BMGS should be interpreted as a structured amplitude amplification process with segmented oracle implementation, rather than as a collection of independent probabilistic segment searches.}\\
Approximating for small $\omega$:
\begin{equation}
P_{\text{BMGS, total}}(t) \approx 1 - e^{-d P_{\text{BMGS}, i}(t)}.    
\end{equation}
Thus, the error probability of our BMGS is:
\begin{equation}
\varepsilon_{\text{BMGS}} = e^{-d P_{\text{BMGS}, i}(t)}.    
\end{equation}

\color{black}
\subsection{Comparison of Success Probability}

The success probability of BMGS follows the same global-amplitude-amplification principle as standard Grover Search. For a search space of size $\mathcal{N} = 2^r$, the Grover rotation angle is $\omega=\arcsin\left(\frac{1}{\sqrt{\mathcal{N}}}\right)$.\\
After $t$ Grover iterations, the exact success probability is governed by $P(t)=\sin^2\left((2t+1)\omega\right)$.\\
Thus, standard Grover Search achieves the maximum success probability when executed for the optimal number of iterations $t^* \approx \frac{\pi}{4}\sqrt{\mathcal{N}}$.\\
Partial Grover Search (PGS) performs amplitude amplification over sub-blocks rather than the full search space. As a result, while it reduces practical search cost, its success probability is generally lower than full Grover Search due to partial amplification and block-level approximation: $P_{\text{PGS}}(t) < P_{\text{GS}}(t)$.\\
In BMGS, segmented oracle operations are combined with bi-directional search across multiple qubit blocks. Although the amplitudes remain globally coupled and are not strictly independent, the structured forward-backward search improves practical convergence and reduces unnecessary amplification over irrelevant regions of the search space.\\
Therefore, for the same effective number of iterations, $P_{\text{BMGS}}(t)
\gtrsim P_{\text{PGS}}(t)$, particularly in shallow segmented implementations where the target structure is efficiently captured.\\
Consequently, the corresponding error probabilities satisfy 
\begin{equation}
\varepsilon_{\text{BMGS}} = 1-P_{\text{BMGS}}(t) < 1-P_{\text{PGS}}(t)
= \varepsilon_{\text{PGS}}.
\end{equation}
Hence, BMGS provides improved practical reliability over PGS while preserving the global Grover amplitude amplification model. Its advantage arises from better circuit efficiency and structured search execution rather than from assuming independent segment-wise probabilities.

\color{black}
\section{Additional Results}
\label{appendx:resulst}

The number of iterations required increased with the number of qubits, with DFGS and BMGS significantly reducing the number of iterations, making them more efficient for larger search spaces, as illustrated in Figure~\ref{fig:iter}. In addition, an increase in segments ($d$) reduces the running time in the worst case, as shown in Figure~\ref{fig:RT_vs_d}. Hence, our BMGS algorithms are optimal for $d=o(\frac{r}{\log_2 b})$ with the constraint $dk < r$, a significant advancement in quantum search algorithms by offering notable reductions in hardware requirements and circuit depth compared to the GS.\\
An advantage of our BMGS algorithm is its segmented approach, which enables parallel processing of each 2-qubit search segment, provided each segment includes a designated auxiliary qubit. This parallelism reduces the number of iterations to 1, substantially improving runtime efficiency compared to the standard Grover search algorithm. The proposed BMGS algorithm's ability to conduct concurrent searches on 2-qubit segments sets it apart. It underscores its potential to revolutionize quantum search methodologies by offering scalability and efficiency in handling multi-solution search scenarios.\\
Due to the segmentation of the entire search space as forward and backward passes, the proposed BMGS algorithm necessitates smaller Oracles over each segment, representing a significant advantage of quantum search with reduced Oracles. As the database size increases, the circuit depth of Oracles escalates, posing challenges for error-free operation.
\setcounter{figure}{2}
\begin{figure}[t] 
\begin{center}
\subcaptionbox{}{\includegraphics[clip, trim=0.0cm 0.0cm 0.0cm 0.0cm, width=0.47\textwidth]{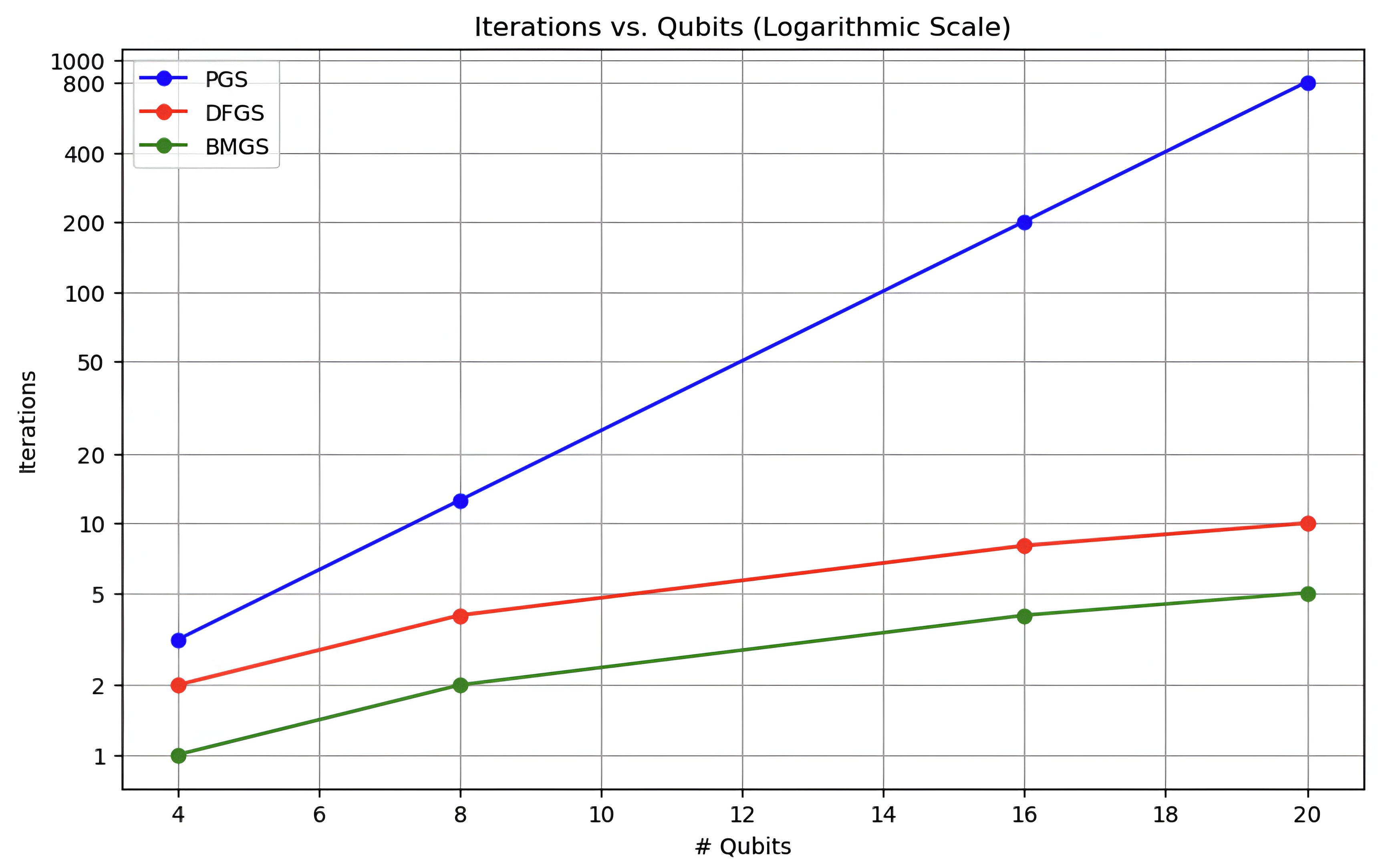}} 
\subcaptionbox{}{\includegraphics[clip, trim=0.0cm 0.0cm 0.0cm 0.0cm, width=0.47\textwidth]{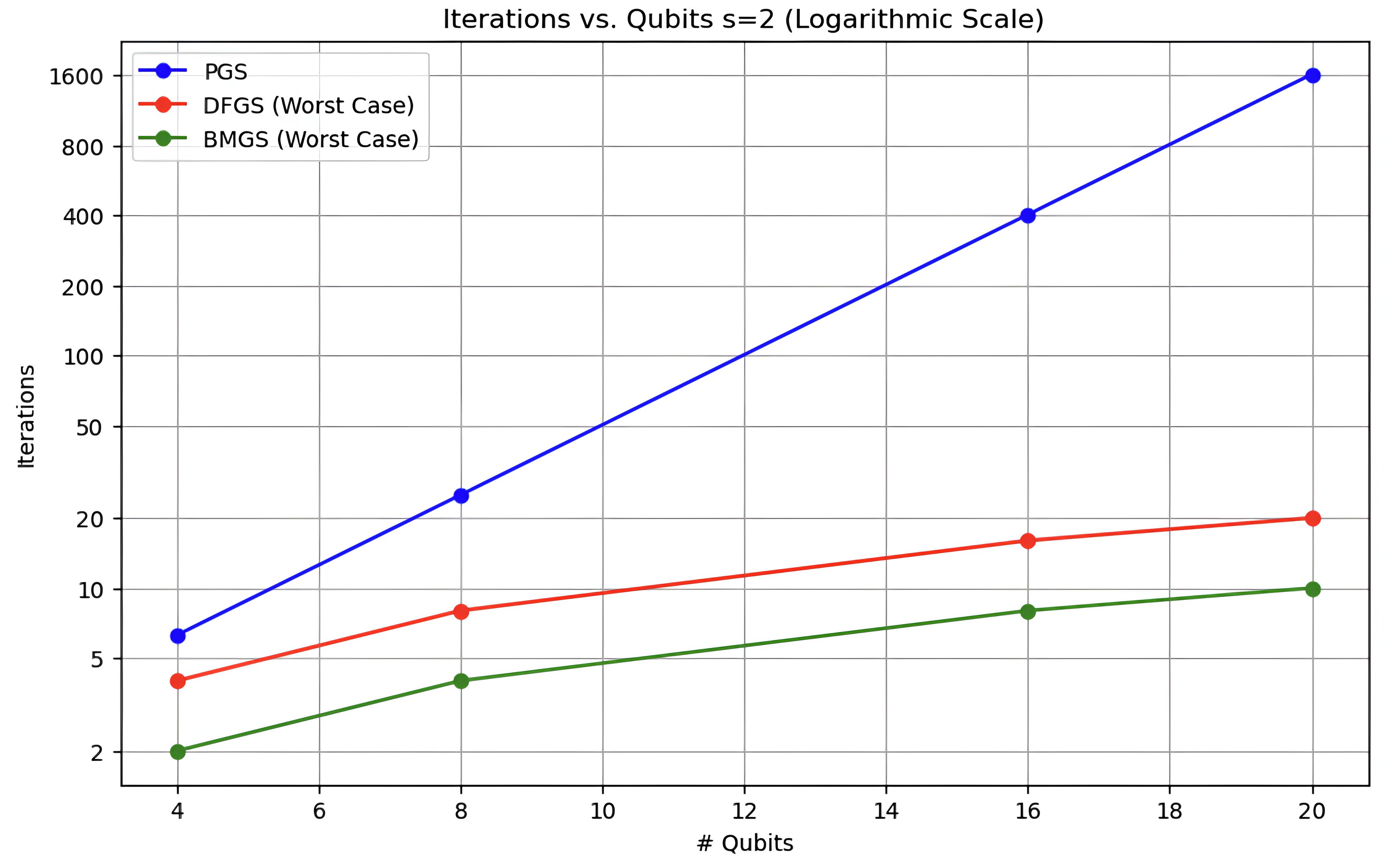}}
\subcaptionbox{}{\includegraphics[clip, trim=0.0cm 0.0cm 0.0cm 0.0cm, width=0.47\textwidth]{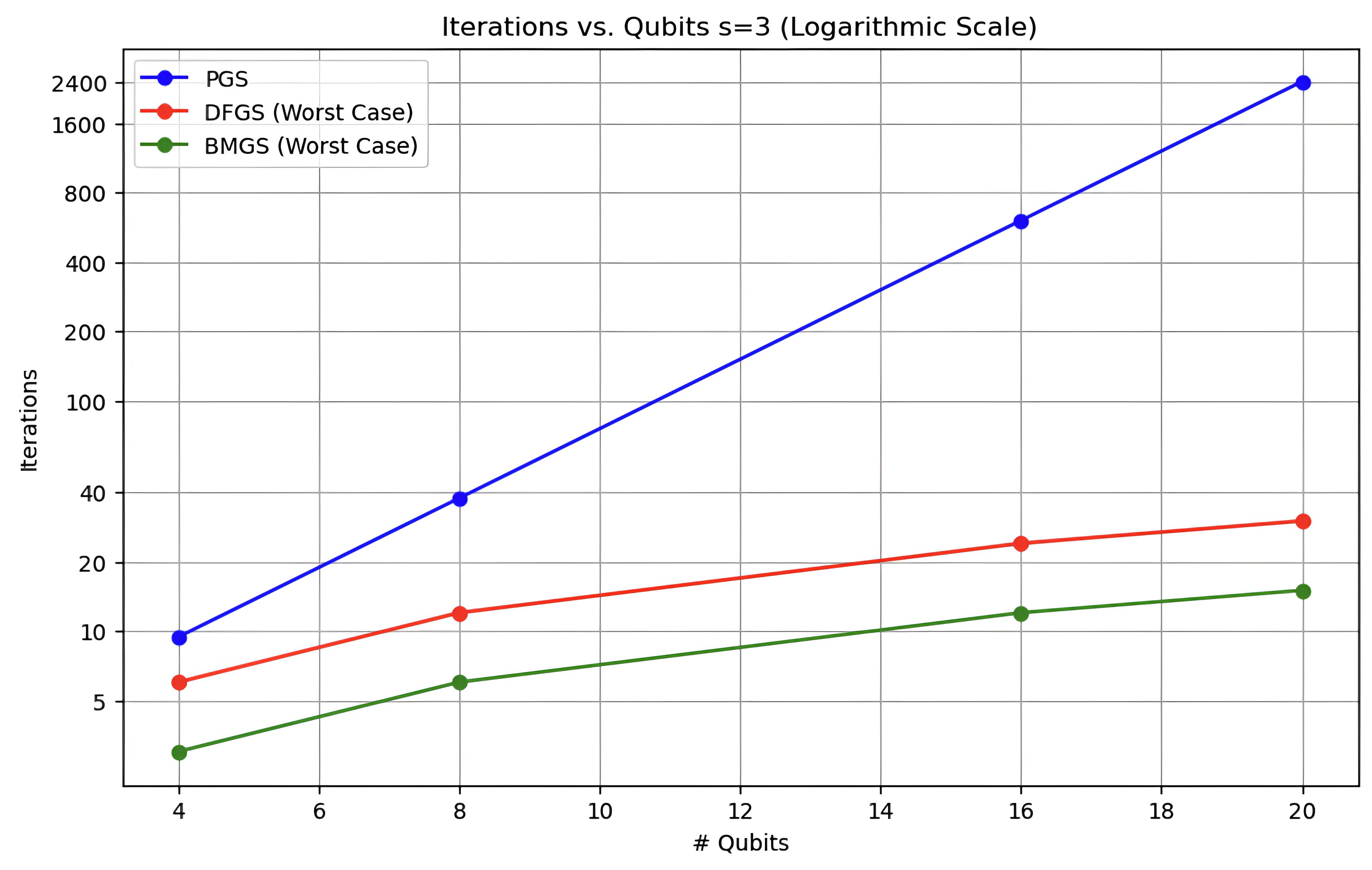}}
\caption{Plot for $\#$Iterations vs. $\#$Qubit for (a) $s=1$, (b) $s=2$, and (c) $s=3$ search space using the proposed BMGS, DFGS~\cite{guo2022}, and PGS~\cite{grover2005}.} 
\label{fig:iter}
\end{center}
\end{figure}

\begin{figure}[t] 
\begin{center}
\subcaptionbox{}{\includegraphics[clip, trim=0.0cm 0.0cm 0.0cm 0.0cm, width=0.47\textwidth]{new_graphs/time_vs_qubits_s=1.jpg}}
\subcaptionbox{}{\includegraphics[clip, trim=0.0cm 0.0cm 0.0cm 0.0cm, width=0.47\textwidth]{new_graphs/time_vs_qubits_s=2.jpg}}
\subcaptionbox{}{\includegraphics[clip, trim=0.0cm 0.0cm 0.0cm 0.0cm, width=0.47\textwidth]{new_graphs/time_vs_qubits_s=3.jpg}}
\caption{Plot for Runtime (s) vs. $d$ for (a) $s=1$, (b) $s=2$, (c) $s=3$ search space using the proposed BMGS.} 
\label{fig:RT_vs_d}
\end{center}
\end{figure}
\subsection{Code availability}
The Qiskit Python implementation of the proposed BMGS algorithm is available on Github: \url{https://anonymous.4open.science/r/Multi-Solution-DFGS-BMGS-B507/}. 

\end{document}